\documentclass[reprint,
 amsmath,amssymb,
aip, jcp
]{revtex4-2}

\pdfoutput=1
\usepackage{makeidx}
\makeindex
\usepackage[]{graphicx}
\usepackage{dcolumn}
\usepackage{bm}
\usepackage[]{hyperref}

\usepackage[
text={7in,10in},centering,
]{geometry}

\usepackage{amssymb}
\usepackage[caption=false]{subfig}
\usepackage{booktabs}
\usepackage{braket}
\usepackage{esint}
\usepackage{multirow}
\usepackage[ngerman,english]{babel}

\newcommand{\myciteauthor}[1]{{\protect\NoHyper\citeauthor{#1}\endNoHyper}}
\newcommand{\uint}{\ointctrclockwise_{\mathcal{U}}}
\DeclareGraphicsExtensions{.pdf,.jpg,.png}

\begin{document}

\widetext{\noindent This article may be downloaded for personal use only. Any other use requires prior permission of the author and AIP Publishing. This article appeared in J. Chem. Phys. 154, 154103 (2021) and may be found at \url{https://doi.org/10.1063/5.0045400}.}

\title{Optimized effective potentials from the random-phase approximation: Accuracy of the quasiparticle approximation}

\author{Stefan Riemelmoser}
\email{stefan.riemelmoser@univie.ac.at}
 \affiliation{%
 Faculty of Physics and Center for Computational Materials Science, University of Vienna, Kolingasse 14-16,
A-1090 Vienna, Austria 
}
\author{Merzuk Kaltak} 
\affiliation{%
VASP Software GmbH, Sensengasse 8/17,
A-1090 Vienna, Austria 
}
\author{Georg Kresse}%
\affiliation{%
 Faculty of Physics and Center for Computational Materials Science, University of Vienna, Kolingasse 14-16,
A-1090 Vienna, Austria 
}%




\date{\today }\vspace{\baselineskip}

\begin{abstract}
The optimized effective potential (OEP) method presents an unambiguous way to construct the Kohn-Sham potential corresponding to a given diagrammatic approximation for the exchange-correlation functional. The OEP from the random-phase approximation (RPA) has played an important role ever since the conception of the OEP formalism. However, the solution of the OEP equation is computationally fairly expensive and has to be done in a self-consistent way. So far, large scale solid state applications have therefore been performed only using the quasiparticle approximation (QPA), neglecting certain dynamical screening effects. We obtain the exact RPA-OEP for 15 semiconductors and insulators by direct solution of the linearized Sham-Schlüter equation. We investigate the accuracy of the QPA on Kohn-Sham band gaps and dielectric constants, and comment on the issue of self-consistency.

\end{abstract}
%
\maketitle
%
%

\section{\label{sec:Riemelmoser2021_1} Introduction}

In Kohn-Sham density functional theory (KS-DFT),\cite{Hohenberg1964,Kohn1965} the complicated many-body problem of interacting electrons in a solid is mapped to a fictitious system of non-interacting particles in an effective local, one-body potential. The mapping is such that the densities of the interacting and non-interacting systems agree, and so do other ground state properties, for example total energies. The effective potential, called KS potential, is uniquely determined from the electronic density  (up to a constant). It consists of the external potential, a classical Coulomb term as well as an ``everything else'' exchange-correlation term, $v_{\rm xc}=\delta E_{\rm xc}/\delta n$. 

The exchange-correlation energy $E_{\rm xc}$ is a generally unknown non-local density functional, and the initial success of KS-DFT was grounded in large part in the accuracy of simple approximations for $E_{\rm xc}$. In the local density approximation (LDA),\cite{Kohn1965} it is approximated locally by that of the homogeneous electron gas with the same density. Later, the LDA has been improved by taking gradient corrections into account, in combination with exact constraints that $E_{\rm xc}$ must fulfill (generalized gradient approximation or GGA).\cite{Perdew1996,*Perdew1997} Although new functionals keep being developed in this manner, see for example Ref. \onlinecite{Yang2016},  systematic improvements within the KS framework are difficult beyond a certain point. 

The search for the exact KS potential, which exhibits interesting properties such as the famous derivative discontinuity,\cite{Perdew1982,Perdew1983} is a worthwhile endeavor in its own right from a theoretical standpoint. Furthermore, high quality KS potentials can serve as a testing ground for lower level approximations. As the KS potential is easily visualizable, the differences can be intuitively interpreted and understood. Last not least, accurate KS potentials provide good starting points for various methods of many-body perturbation theory (MBPT). 

In principle, quasi-exact KS potentials can be constructed from accurate densities obtained via quantum chemistry methods, as was done, for example, by \myciteauthor{Umrigar1994}.\cite{Umrigar1994} This approach is, however, limited to very small systems due to its unfavorable scaling with respect to system size.
Alternatively, it is also possible to take an energy functional from MBPT and map it to a KS potential. Through this procedure, which is known as the optimized effective potential (OEP) method, ever better approximations for $v_{\rm xc}$ can be found by including higher order terms in the perturbation series. 

The OEP idea dates back to the 1950s, when \citet{Slater1951} constructed a local approximation for the Hartree-Fock potential by taking an orbital average. After further development of the exchange-only theory,\cite{Sharp1953,Talman1976} the framework was later generalized by \citet{Sham1983} and others. \cite{Sham1985,Casida1995,Barth2005} 
Already in the 1980s, Godby \textit{et al.}\cite{Godby1986,Godby1988} pioneered the application of OEP from the random-phase approximation (RPA-OEP). They were able to prove that the derivative discontinuity plays a large role for the band gaps of semiconductors, which had been debated up to this point. Their findings were later confirmed by \myciteauthor{Gruening2006}\cite{Gruening2006,Gruening2006a} and extended to wide-gap systems. More recently, the RPA-OEP for small molecules was extensively studied by Hellgren \textit{et al.},\cite{Hellgren2007,Hellgren2012,Caruso2013,Hellgren2015} who have also investigated improvements beyond the RPA via the time-dependent DFT framework.\cite{Hellgren2008,Hellgren2010}
Similar studies were also performed by Bleiziffer \textit{et al.}\cite{Bleiziffer2013,Bleiziffer2013a,Bleiziffer2015} 
The RPA-OEP exhibits several features that are expected from the exact exchange-correlation potential but are missing in the (semi-)local approximations. For example, it correctly decays as $1/r$ far away from finite systems such as atoms.\cite{Sham1985,Engel2003} Moreover, it yields a good description of atomic shell oscillations\cite{Hellgren2007} as well of the dissociation limit of closed-shell molecules.\cite{Hellgren2012}

Due to its prohibitive computational cost, applications to the solid state have so far been limited to few systems. The most comprehensive study of the RPA-OEP for solids up to date was done by \myciteauthor{Klimes2014}.\cite{Klimes2014}  However, they did use a static approximation for the self-energy, the so-called quasiparticle approximation (QPA). The main focus of the current work is to drop this approximation in order to obtain KS potentials from self-consistent RPA-OEP. These results then serve as reference to determine the accuracy of the QPA.

The paper is organized as follows. In Sec. \ref{sec:Riemelmoser2021_2}, we briefly recapitulate the OEP method and give the RPA-OEP expressions. The QPA is introduced, and we discuss its effects for simple model systems. In Sec. \ref{sec:Riemelmoser2021_3}, key details of our implementation are presented.	In Sec. \ref{sec:Riemelmoser2021_4}, we show results from RPA-OEP calculations for 15 semiconductors and insulators. The effect of the QPA on band gaps and dielectric constants is investigated. The results are discussed in Sec. \ref{sec:Riemelmoser2021_5}, and conclusions are drawn in Sec.	\ref{sec:Riemelmoser2021_6}. The current work is restricted to the case of non-spin-polarized electrons, spin is accounted for by inclusion of factors of two. Unless stated otherwise, Hartree units are used throughout the work.

\section{\label{sec:Riemelmoser2021_2} Theory}
\subsection{The optimized effective potential method}

\begin{figure}[!!tb]
\centering
\includegraphics [width=\linewidth,keepaspectratio=true] {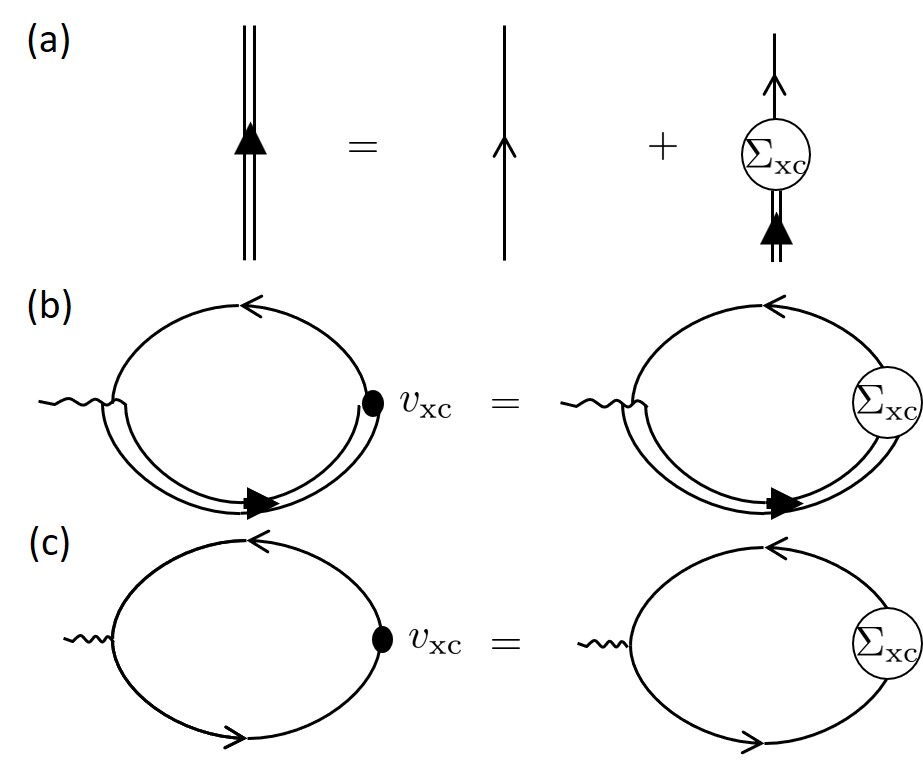}
\caption{Diagrammatic representation\cite{Fetter2003,Sham1985} of (a) the Dyson equation, (b) the Sham-Schlüter equation, and (c) the linearized Sham-Schlüter equation. The self-energy $\Sigma_{\rm xc}$ is replaced by a local potential $v_{\rm xc}$ such that the density of the interacting system $n$ is reproduced exactly (b), or up to first order in $\Sigma_{\rm xc}$ respectively (c).}
\label{fig:Sham-Schlueter}
\end{figure}

We begin our discussion with a brief derivation of the Sham-Schlüter equation (SSE),\cite{Sham1983,Engel2003} which provides the theoretical basis for the OEP method.  The Dyson equation connects the interacting Green's function $G$ to the non-interacting (Kohn-Sham) Green's function $G_0$ and the (exchange-correlation part of the) self-energy $\Sigma_{\rm xc}$ \cite{Fetter2003}
\begin{equation}
\begin{aligned}
&G(\textbf{r}_2,\textbf{r}_1,\omega) = G_0(\textbf{r}_2,\textbf{r}_1,\omega)\\ + &\int \text{d}\textbf{r}_3\int \text{d}\textbf{r}_4	G_0(\textbf{r}_2,\textbf{r}_3,\omega)\Sigma_{\rm xc}(\textbf{r}_3,\textbf{r}_4,\omega) G(\textbf{r}_4,\textbf{r}_1,\omega),
\end{aligned}
\end{equation}
where the representation in terms of frequencies $\omega$ assumes time translation invariance. By closing the loops, i.e. by setting $\textbf{r}_2=\textbf{r}_1$, summing over spin and integrating over frequencies, we obtain
\begin{equation}
\begin{aligned}
&\underbrace{2\uint \frac{\text{d}\omega}{2\pi i} G(\textbf{r}_1,\textbf{r}_1,\omega)}_{n(\textbf{r}_1)} = \underbrace{2\uint \frac{\text{d}\omega}{2\pi i}G_0(\textbf{r}_1,\textbf{r}_1,\omega)}_{n_0(\textbf{r}_1)} \\&+ 2\uint \frac{\text{d}\omega}{2\pi i}G_0(\textbf{r}_1,\textbf{r}_3,\omega)\Sigma_{\rm xc}(\textbf{r}_3,\textbf{r}_4,\omega)G(\textbf{r}_4,\textbf{r}_1,\omega) .
\end{aligned}
\end{equation}
Here, $n$ and $n_0$ are the densities of the interacting system and non-interacting system, respectively, and $\uint$ denotes a counterclockwise contour about the upper complex half plane. Now we replace $\Sigma_{\rm xc}$ by a local, frequency independent potential $v_{\rm xc}$, and require that this reproduces $n$ exactly,
\begin{widetext}
\begin{equation}\label{eq:SSE}
\begin{aligned}
n(\textbf{r}_1)- n_0(\textbf{r}_1)
&= 2\int \text{d}\textbf{r}_3\int \text{d}\textbf{r}_4 \uint \frac{\text{d}\omega}{2\pi i} G_0(\textbf{r}_1,\textbf{r}_3,\omega)\Sigma_{\rm xc}(\textbf{r}_3,\textbf{r}_4,\omega)G(\textbf{r}_4,\textbf{r}_1,\omega)  \\
&\overset{!}{=} 2\int \text{d}\textbf{r}_3\int \text{d}\textbf{r}_4\uint\frac{\text{d}\omega}{2\pi i}G_0(\textbf{r}_1,\textbf{r}_3,\omega) v_{\rm xc}(\textbf{r}_3)\delta(\textbf{r}_3-\textbf{r}_4)G(\textbf{r}_4,\textbf{r}_1,\omega) .
\end{aligned}
\end{equation}
\end{widetext}
This result, the SSE, is depicted diagrammatically in Fig. \ref{fig:Sham-Schlueter}. By virtue of the Hohenberg-Kohn theorems,\cite{Hohenberg1964} $v_{\rm xc}$ is, in fact, just the exact KS exchange-correlation potential, if no approximations for $\Sigma_{\rm xc}$ are made. Hence, $v_{\rm xc}$ is uniquely defined by the SSE (up to a constant). Furthermore, the ambiguity with respect to a constant shift can be lifted by the fact that the DFT chemical potential is correct. \cite{Godby1988}

In practice, the OEP is often approximated by solving the linearized SSE, \cite{Godby1988}
\begin{equation}\label{eq:LSSE}
\begin{aligned}
\int \text{d} \textbf{r}' \chi_0(\textbf{r},\textbf{r}') v_{\rm xc}(\textbf{r}') = & 2\int \text{d}\textbf{r}' \int \text{d}\textbf{r}'' \uint\frac{\text{d}\omega}{2\pi i} G_0(\textbf{r},\textbf{r}',\omega) \\
&\times  \Sigma_{\rm xc}(\textbf{r}',\textbf{r}'',\omega) G_0(\textbf{r}'',\textbf{r},\omega) ,
\end{aligned}
\end{equation}
which is obtained by replacing $G \rightarrow G_0$ in Eq. \eqref{eq:SSE}, and identifying the frequency integral in the last line as the static response function of the non-interacting system, symbolically $\chi_0=G_0G_0$.
Since Eq. \eqref{eq:LSSE} is linear in $\Sigma_{\rm xc}$, we can decompose $v_{\rm xc}$ into separate exchange and correlation parts
\begin{equation}
\begin{aligned}
&\Sigma_{\rm xc} = \Sigma_{\rm x} + \Sigma_{\rm c} \\
& v_{\rm xc}[\Sigma_{\rm xc}] = v_{\rm x}[\Sigma_{\rm x}]+v_{\rm c}[\Sigma_{\rm c}] .
\end{aligned}
\end{equation}
For alternative derivations and more in-depth discussion of the OEP method, we refer to the reviews of \citet{Engel2003} and \myciteauthor{Kuemmel2008}.\cite{Kuemmel2008}

\begin{figure}[!!tb]
\centering
\includegraphics [width=\linewidth,keepaspectratio=true] {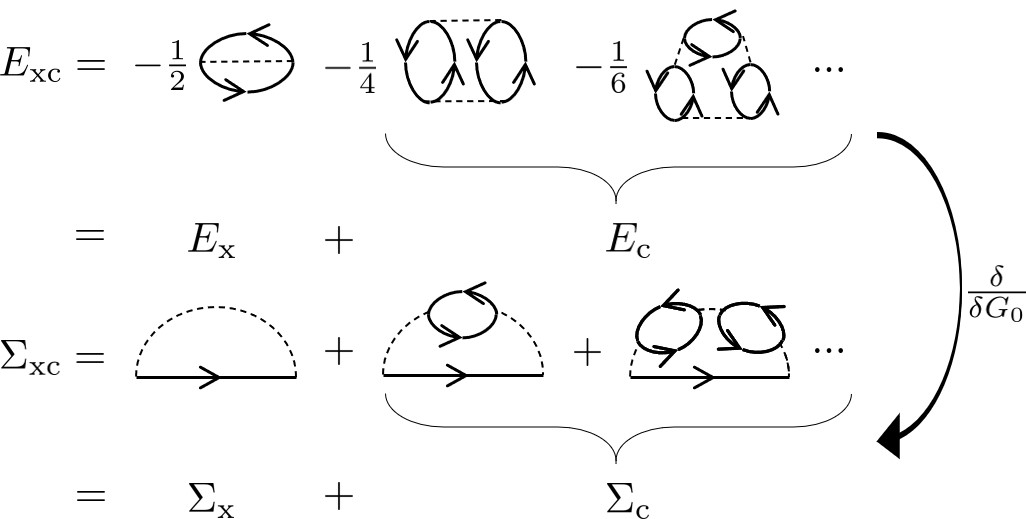}
\caption{Diagrammatic representation of the RPA exchange-correlation energy.  The self-energy in the $G_0W_0$ approximation is obtained by taking the functional derivative with respect to the non-interacting Green's function, $\delta E_{\rm xc}/\delta G_0 = \Sigma_{\rm xc}$.\cite{Hellgren2007}}
\label{fig:GW_diagrams}
\end{figure}

We now turn to specific approximations for the self-energy $\Sigma_{\rm xc}$. The exchange-only approximation  (EXX-OEP),
\begin{equation}
\Sigma_{\rm x}(\textbf{r},\textbf{r}') = -\sum_i^{\rm occ}\frac{\phi_i(\textbf{r})\phi^*_i(\textbf{r}')}{|\textbf{r}-\textbf{r}'|},
\end{equation}
is the prototypical static approximation. Here, the frequency integral on the RHS of equation \eqref{eq:LSSE} can be performed analytically yielding
\begin{equation}\label{eq:EXX-OEP}
\begin{aligned}
&\int \text{d} \textbf{r}_3 \chi_0(\textbf{r},\textbf{r}') v_{\rm x}(\textbf{r}') = 2\sum_i^{\rm occ}\sum_a^{\rm unocc}\int \text{d}\textbf{r}' \int \text{d}\textbf{r}'' \\
&\times \phi_a(\textbf{r})\phi_a^*(\textbf{r}') \Sigma_{\rm x}(\textbf{r}',\textbf{r}'') \phi_i(\textbf{r}'')\phi_i^*(\textbf{r})\frac{1}{\varepsilon_i-\varepsilon_a} + {\rm c.c.}, 
\end{aligned}
\end{equation}
where the $\phi_i$ and $\varepsilon_i$ denote the orbitals of the non-interacting system and corresponding eigenvalues respectively. In RPA-OEP, the $GW$ approximation\cite{Hedin1965,Hedin1970,Hybertsen1986} (more specifically, its one-shot or $G_0W_0$ version) for the self-energy is used,
\begin{equation}
\begin{aligned}
\Sigma_{\rm xc}(\textbf{r},\textbf{r}',\omega) 
=
- \uint \frac{\text{d}\omega'}{2\pi i} G_0(\textbf{r},\textbf{r}',\omega+\omega')W_0(\textbf{r},\textbf{r}',\omega') .
\end{aligned}
\end{equation}
Here, $W_0$ is the screened Coulomb interaction using the random-phase approximation for the polarizability,
\begin{equation}\label{eq:RPA_screening}
W_0 = V + \chi_0V\chi_0 + \chi_0V\chi_0V\chi_0 + ... = V(1-\chi_0V)^{-1} ,
\end{equation}
where $V$ is the bare Coulomb potential.  Fig. \ref{fig:GW_diagrams} shows the diagrammatic representation of these two approximations for the self-energy. In an orbital basis, the RPA-OEP equation reads
\begin{equation}\label{eq:RPA_OEP}
\begin{aligned}
&2\sum_i^{\rm occ}\sum_a^{\rm unocc} \frac{\phi_i(\textbf{r})\phi_a^*(\textbf{r}) v_{\text{xc}, ia}}{\varepsilon_i-\varepsilon_a} + \rm{c.c.} = \\&\sum_{mn}^{\rm all} \phi_m(\textbf{r})\phi_n^*(\textbf{r})\rho_{\text{xc},mn} + \rm{c.c.}\\
&\rho_{\text{xc},mn} = 2\sum_{rs}^{\rm{all}} \uint \frac{\text{d}\omega}{2\pi i} G_{0,mr}(\omega)\Sigma_{\text{xc},rs}(\omega)G_{0,sn}(\omega) ,
\end{aligned}
\end{equation}
where the respective matrix elements for any operator $A$ are given by $A_{mn}=\braket{m|A|n}$. It is interesting to note that the RPA density matrix, $n_{\rm xc}(\textbf{r},\textbf{r}') = \sum_{mn}\phi_m(\textbf{r})\phi^*_n(\textbf{r}')\rho_{\text{xc},mn}$,  is inherently connected to the theory of RPA natural orbitals. That is, the matrix $\rho_{\text{xc},mn}$ is per definition diagonal in the natural orbital basis, see Ref. \onlinecite{Ramberger2019} for further discussion. We also use the natural orbital basis in the present RPA-OEP implementation (compare Sec. \ref{sec:Riemelmoser2021_3}). 

Let us summarize briefly what we have achieved so far. We start from a MBPT energy functional that we want to map to an optimized effective potential (in our case exact exchange and RPA, respectively). An approximate MBPT functional corresponds to a subset of Feynman diagrams. Its variation with respect to the non-interacting Green's function yields a diagrammatic approximation for the self-energy (in our case the exchange-only and $G_0W_0$ approximation, respectively, compare also Fig. \ref{fig:GW_diagrams}).  The self-energy is then inserted into the linearized SSE, yielding the OEP. This OEP truly is the Kohn-Sham potential corresponding to the given energy functional. For a more rigorous discussion of this formalism, we refer to \myciteauthor{Hellgren2007}.\cite{Hellgren2007}

Historically, the RPA has been brought into the DFT framework via the adiabatic-connection formalism.\cite{Langreth1975,Langreth1977} The expression for the exchange-correlation energy obtained in this way is formally equivalent to the diagrammatic expansion shown in Fig. \ref{fig:GW_diagrams}. Following a general procedure,\cite{Sham1983,Goerling1994} it is also possible to re-derive the RPA-OEP equation \eqref{eq:RPA_OEP}	by taking the functional derivative, $v_{\rm xc}(\textbf{r})=\delta E_{\rm xc}/\delta n(\textbf{r})$, and applying the chain rule (see Ref. \onlinecite{Niquet2003}). 

\subsection{Quasiparticle approximation}

Now that we have established the RPA-OEP framework, we discuss in the following an additional approximation, which is central to the current work. 

The goal of the QPA, as first proposed by \myciteauthor{Casida1995},\cite{Casida1995} is to find a static approximation $\Sigma_{\rm xc}(\omega) \approx \Sigma_{\rm xc}^{\rm QPA}$ for the self-energy.  In the QP limit, we can assume that $\Sigma_{\rm xc}(\omega)$ is a slowly varying function of $\omega$, and evaluate it at the QP energies $\varepsilon_m$. This simplifies the RPA-OEP equation \eqref{eq:RPA_OEP}, as one can perform the frequency integral on the RHS as in the exchange-only case. The resulting equation in an orbital basis is\cite{Casida1995}
\begin{equation}\label{eq:QPA_RPA_OEP}
\begin{aligned}
&2\sum_{i}^{\rm occ. }\sum_a^{\rm unocc.} \phi_i(\textbf{r}) \phi_a^*(\textbf{r}) \frac{v_{\text{xc},ia}}{\varepsilon_i-\varepsilon_a} + \rm{c.c.} = \\
&2\sum_{i}^{\rm occ. }\sum_a^{\rm unocc.} \phi_i(\textbf{r}) \phi_a^*(\textbf{r}) \frac{\Sigma^{\rm QPA}_{\text{xc},ia}}{\varepsilon_i-\varepsilon_a} + \rm{c.c.} 
\end{aligned}
\end{equation}
This is just the EXX-OEP equation with $\Sigma_{\rm x} \rightarrow \Sigma_{\rm xc}^{\rm QPA}$, compare Eq. \eqref{eq:EXX-OEP}. There is some ambiguity about how one should choose the non-diagonal elements of $\Sigma_{\rm xc}^{\rm QPA}$, which account for non-local effects. \cite{Casida1995,Faleev2004,Schilfgaarde2006,Kotani2007} Here, we follow \citet{Klimes2014} and choose the Hermitian form\cite{Schilfgaarde2006}

\begin{equation}\label{eq:Kotani}
\Sigma_{\text{xc}, mn}^{\rm QPA} = \frac{1}{2} \left[\text{Re}\Sigma_{\text{xc},mn} (\varepsilon_m) + \text{Re}\Sigma_{\text{xc},mn} (\varepsilon_n)\right] .
\end{equation}

It is interesting to note that Kotani \textit{et al.}\cite{Faleev2004,Schilfgaarde2006,Kotani2007,Kotani2007a} used the QPA not as an approximation for the RPA-OEP potential, but rather to obtain a static, non-local potential,
\begin{equation}\label{eq:Kotani_nl}
V_{\rm xc}^{\rm nl} = \sum_{mn}^{\rm all}\ket{m}\Sigma_{{\rm xc},mn}^{\rm QPA}\bra{n}, 
\end{equation}
for self-consistent $GW$ calculations. This approach is similar to the generalized OEP method of \myciteauthor{Jin2017},\cite{Jin2017}
a detailed comparison of these methods is, however, beyond the scope of this paper.
Here, we note only that if one neglects the off-diagonal elements completely, one obtains a local potential as given in Eq. (3.14) of Ref. \onlinecite{Casida1995}, which is a generalized form of the Slater potential. \cite{Slater1951} 

Similar to the potential given in Eq. \eqref{eq:Kotani_nl}, a non-local, static approximation to the $GW$ self-energy can be obtained by Hedin's COHSEX \cite{Hedin1965} (static Coulomb hole plus screened exchange). Like the QPA, COHSEX has been used as a non-local potential for self-consistent $GW$, achieving similar results.\cite{Bruneval2006,Bruneval2014} The difference between those two approximations is that in the QPA, the dynamical effects are neglected at the level of $\Sigma_{\rm xc}$, whereas COHSEX neglects them already at the level of $W$.

\subsection{The QPA in simple model systems}\label{sec:model_QPA}

In order to shed some light on the physics behind the QPA, we investigate its effects in simple model systems. First, we calculate exchange-correlation energies for the homogeneous electron gas (HEG), and then we consider the effect of the QPA on the band gap of a simple flat band model.

For the HEG, spatial symmetry dictates that the OEP is a constant $v_{\rm xc,HEG}$, which depends only on the density of the HEG $n$ or equivalently the Wigner-Seitz radius $r_{\rm s}=(3/4\pi n)^{1/3}$. Furthermore, it can be seen from the linearized SSE that $v_{\rm xc}$ is simply given by the self-energy evaluated at the Fermi edge\cite{Sham1985} 
\begin{equation}\label{eq:Sham_HEG}
v_{\rm xc,\rm HEG} = \Sigma_{\rm xc,HEG}(k_{\rm F},k_{\rm F}^2/2) ,
\end{equation}
where $k_{\rm F}=(\alpha r_{\rm s})^{-1}$ is the Fermi wave vector of the HEG, with $\alpha=(4/9\pi)^{1/3}$.
Any exchange-correlation potential for the HEG is related to the exchange-correlation energy per particle $\varepsilon_{\rm xc,HEG}$ via the general relation 
\begin{equation}\label{eq:standard_correlation}
v_{\rm xc,HEG} = \varepsilon_{\rm xc,HEG} - \frac{r_{\rm s}}{3}\frac{\text{d} \varepsilon_{\rm xc,HEG}}{\text{d} r_{\rm s}} .
\end{equation}
For the exchange-only self-energy, one obtains the familiar Dirac expression $\varepsilon_{\rm x,HEG} = -3/(4\pi\alpha r_{\rm s})$, whereas the correlation part of the $G_0W_0$ energy is connected to the RPA correlation energy\cite{Caruso2013}
\begin{equation}
v_{\rm c,HEG}^{G_0W_0} = \varepsilon_{\rm c,HEG}^{\rm RPA} - \frac{r_{\rm s}}{3}\frac{\text{d} \varepsilon_{\rm c,HEG}^{\rm RPA}}{\text{d} r_{\rm s}} .
\end{equation}
We have used explicit labels here to emphasize that the random-phase approximation is constructed from the \textit{non-interacting} quantities $G_0$ and $W_0$. For the HEG, the self-energy is diagonal in the momentum basis, hence $\Sigma_{\text{xc,HEG}}^{\rm QPA}(k)=\Sigma_{\rm xc,HEG}(k,\varepsilon_k)$. It is evident that the QPA gives the correct OEP per definition. In comparison, the COHSEX yields an OEP which is too negative,\cite{Hedin1965} see also Appendix \ref{App:Riemelmoser2021_A}.

This exact relation for the OEP hides in some sense the drastic simplification that is made, when the self-energy as a function of frequency is crudely approximated as a constant. In reality, is has pronounced structure even for the HEG, especially near the plasmon frequency, see e.g. \myciteauthor{Hedin1970}.\cite{Hedin1970} Any static approximation corresponds to setting the renormalization factor $Z=(1-\partial \text{Re}\Sigma/\partial \omega)^{-1}$ equal to 1. Clearly, all information about lifetimes and satellites is lost in this way.  To demonstrate how the QPA can fail also for total energies, we have calculated exchange-correlation energies via the Galitskii-Migdal (GM) formalism,\cite{Fetter2003} that is we calculate the GM energy per particle\cite{Holm1998}
\begin{equation}\label{eq:Galitskii}
\begin{aligned}
\varepsilon_{\rm HEG}^{\rm GM} &=
\frac{1}{n}\int_0^{\infty}\frac{\text{d}k}{(2\pi)^3}4\pi k^2\uint \frac{\text{d}\omega}{2\pi i}\\ &\times\left[2\varepsilon(k)G(k,\omega)+ G(k,\omega)\Sigma_{\rm xc}(k,\omega)\right],
\end{aligned}
\end{equation}
where $\varepsilon(k)=k^2/2$ is the free electron parabola. The exchange-correlation energy per particle $\varepsilon_{\rm xc, HEG}^{\rm GM}$ is then defined as the difference between the above expression for the interacting and non-interacting HEG, i.e.

\begin{equation}\label{eq:Galitskii_correlation}
\varepsilon_{\rm xc, HEG}^{\rm GM} = \varepsilon_{\rm HEG}^{\rm GM}-3k_{\rm F}^2/10 .
\end{equation}
  
\begin{figure}[!!tb]
\centering
\includegraphics [angle=-90,width=\linewidth,keepaspectratio=true] {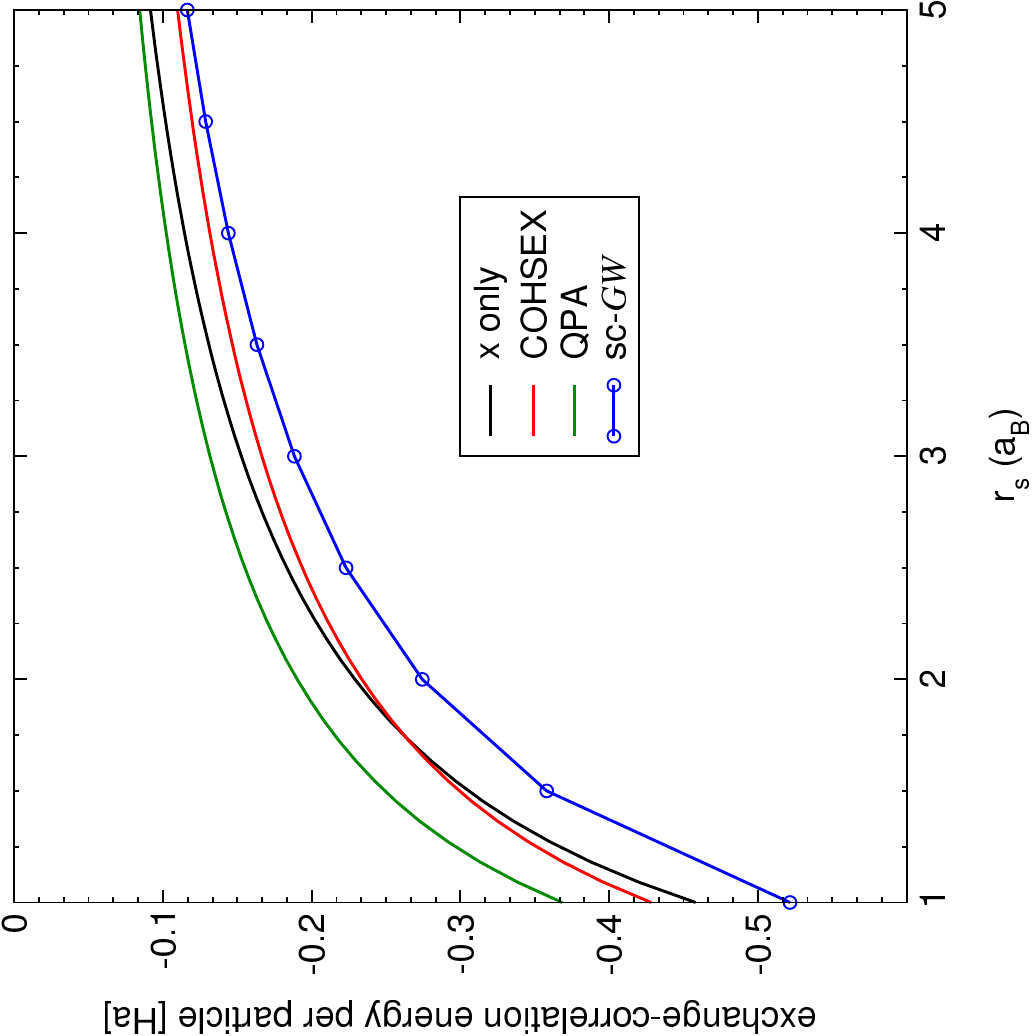}
\caption{Exchange-correlation energies per particle for the HEG via the Galitskii-Migdal formula \eqref{eq:Galitskii_correlation}. Shown are different static approximations for the $GW$ self-energy (x only, COHSEX, QPA) as well as the fully self-consistent case. For further discussion, see also Appendix \ref{App:Riemelmoser2021_A}.}
\label{fig:GW_correlation_energies}
\end{figure}

In general, Eq. \eqref{eq:Galitskii} has to be evaluated self-consistently, where $G$, $W$ and $\Sigma_{\rm xc}$ are updated via Hedin's $GW$ equations. Only then do both expressions for the exchange-correlation energies, Eqs. \eqref{eq:Galitskii_correlation} and \eqref{eq:standard_correlation}, coincide\cite{Holm1999} [in Eq. \eqref{eq:standard_correlation}, $\varepsilon_{\text{xc,HEG}}$ is implicitly defined via the solution of the differential equation, see \citet{Hedin1965}].  For static approximations to the $GW$ self-energy, $\Sigma_{\rm xc}(\omega)\rightarrow \Sigma_{\rm xc}^{\rm static}$, there is no real self-consistency problem. The frequency integral in Eq. \eqref{eq:Galitskii} can be performed analytically, and one obtains the simple expression (see chap. 10 of Ref. \onlinecite{Fetter2003})
\begin{equation}\label{eq:GW_correlation_energy_static}
\varepsilon_{\rm xc,HEG}^{\rm static} =  \frac{1}{n}\int_0^{k_{\rm F}} \frac{\text{d}k}{(2\pi)^3} 4\pi k^2 \Sigma_{\rm xc, HEG}^{\rm static}(k) .
\end{equation}
This result can be intuitively understood, as a static interaction can only shift, but never smear the Fermi surface, since the QP lifetime is always infinite. For example, in the familiar exchange-only case, one obtains\cite{Hedin1970}
\begin{equation}\label{eq:Sigma_EXX}
\begin{aligned}
&\Sigma_{\rm x}(k) = \frac{k_{\rm F}}{\pi} \left(1+\frac{k_{\rm F}^2-k^2}{2kk_{\rm F}}\ln\left|\frac{k_{\rm F}+k}{k_{\rm F}-k}\right|\right),
\end{aligned}
\end{equation}
which yields once again the Dirac expression when inserted into Eq. \eqref{eq:GW_correlation_energy_static}.
We have performed numerical calculations for $\Sigma_{\rm xc}$ in the COHSEX and QPA as well as in the fully-self consistent case (sc-$GW$).  
The procedure is detailed further in Appendix \ref{App:Riemelmoser2021_A}, results are depicted in Fig. \ref{fig:GW_correlation_energies}.

It can be seen that both static approximations, COHSEX and QPA, yield GM exchange-correlation energies similar to the exchange-only case, that is, too positive when compared to sc-$GW$. We want to stress that the latter is indeed the most sound from a physical stand point, and any deviance must be interpreted as a failure of the static approximations. Furthermore, it was noted previously by \myciteauthor{Holm1998}\cite{Holm1998,Holm1999} that sc-$GW$ is very close to the exact (QMC) data. Here, we find that the QPA performs even worse than COHSEX, which at least correctly predicts a negative contribution from correlation for metallic densities. As we discuss in Appendix \ref{App:Riemelmoser2021_A}, the main failure of the QPA is that no QP weight is transferred from the QP peak to lower lying excitations (satellites).

After having discussed the QPA for the HEG, we now estimate its effect on the band gap of a semiconductor. To this purpose, we construct a simple flat band model, which allows us to derive a first-order estimate for the band-gap correction $\Delta E_{\rm g}$, i.e. the difference 
\begin{equation}\label{eq:delta_g}
\Delta E_{\rm g} = E_{\rm g}^{\rm QPA} - E_{\rm g}, 
\end{equation}
where $E_\text{g}$ is the RPA-OEP band gap. This procedure is inspired by the work of \myciteauthor{Hanke1988},\cite{Hanke1988} who sought to build an analytical model for the $G_0W_0$ self-energy and the corresponding $v_{\rm xc}$ from RPA-OEP. 

The assumptions of our model are as follows: (i) We approximate the system by two flat bands, one valence band and one conduction band (extreme tight-binding approximation). The bands have energies $\varepsilon_v$ and $\varepsilon_c$ respectively, and the OEP band gap is $E_{\text{g}}=\varepsilon_c-\varepsilon_v$. Furthermore, the non-interacting Green's function is given by
\begin{equation}
\begin{aligned}
&G_{0}(\textbf{r},\textbf{r}',\omega) = \frac{\phi_v(\textbf{r})\phi^*_v(\textbf{r}')}{\omega-\varepsilon_v-i\eta}
+\frac{\phi_c(\textbf{r})\phi^*_c(\textbf{r}')}{\omega-\varepsilon_c+i\eta},
\end{aligned}
\end{equation}
where $\eta$ is a positive infinitesimal.

(ii) We expand the self-energy as a function of frequency linearly around the QP energies, neglecting for a moment any energy dependence of the off-diagonal terms,
\begin{equation}
\begin{aligned}\label{eq:self_energy_first_order}
&\Sigma_{\text{xc},vv}(\omega) = \Sigma_{\text{xc},vv}(\varepsilon_{v}) + (\omega-\varepsilon_{v})\Sigma'_{vv}(\varepsilon_v) \\
&\Sigma_{\text{xc}, cc}(\omega) = \Sigma_{\text{xc},cc}(\varepsilon_{c}) + (\omega-\varepsilon_{c})\Sigma'_{cc}(\varepsilon_c) \\
&\Sigma_{\text{xc},vc} = \Sigma_{\text{xc},cv}^* = \Sigma^{\rm QPA}_{\text{xc},vc},
\end{aligned}
\end{equation}
where $\Sigma'$ is connected to the $Z$-factor by 
\begin{equation}\label{eq:Z_factor}
Z= \frac{1}{1-\Sigma'},
\end{equation}
and is therefore negative as $Z<1$. The QPA can be obtained by setting $\Sigma' \rightarrow 0$ in Eq. \eqref{eq:self_energy_first_order} and in the following expressions. 

The second assumption implies that the effect of the QPA will be small in the sense $\Delta E_{\rm g}/E_{\rm g} \ll 1$. Within the limits of this approximation, we can solve the linear SSE in a one-shot fashion, neglecting the self-consistency requirement. Inserting assumption (i) in the OEP equation \eqref{eq:RPA_OEP} and performing the contour integrals yields
\begin{equation}
\begin{aligned}
&-2\frac{\phi_v(\textbf{r})\phi^*_c(\textbf{r})v_{\text{xc},vc}}{E_{\rm g}} + \text{c.c.}=\\
\bigg[&-2\frac{\phi_v(\textbf{r})\phi^*_c(\textbf{r})\Sigma^{\rm QPA}_{\text{xc},vc}}{E_{\rm g}} +2\phi_v(\textbf{r})\phi^*_v(\textbf{r})\Sigma'_{vv}(\varepsilon_v) \bigg] +  \text{c.c.} 
\end{aligned}
\end{equation}
Comparing this result with Eq. \eqref{eq:QPA_RPA_OEP}, we see that only the pole in $\Sigma_{\text{xc},vv}$ yields an extra term on the RHS, proportional to $E_{\rm g}\Sigma'_{vv}$.	 By inserting the completeness relation $\ket{v}\bra{v}+\ket{c}\bra{c}=1$, we find
\begin{equation}
\Delta v_{\rm xc}(\textbf{r}) = \Delta v_{\text{xc},vv}+E_{\text{g}}\Sigma'_{vv} , 
\end{equation}
where we have defined $\Delta v_{\rm xc}=v_{\rm xc}^{\rm QPA}-v_{\rm xc}$. As $\Delta v_{\rm xc}(\textbf{r})$ is constant, we see that here, the QPA amounts only to a rigid shift of both bands. Turning now to the influence of the off-diagonal terms [compare Eq. \eqref{eq:Kotani}], and proceeding as above, we find that valence and conduction band are further shifted by
\begin{equation}
\begin{aligned}
&\Delta v_{\text{xc},vv} = \frac{E_\text{g}\Sigma'_{vv}(\varepsilon_v)}{2}\\
&\Delta v_{\text{xc},cc} = \frac{E_\text{g}\Sigma'_{cc}(\varepsilon_c)}{2} ,
\end{aligned}
\end{equation}
where we have used that $\Sigma'(\varepsilon_v)\approx \Sigma'(\varepsilon_c)$ within the linear approximation. Using Eq. \eqref{eq:Z_factor} to replace $\Sigma'$ by the $Z$-factor, we obtain
\begin{equation} \label{eq:flat_band}
\Delta E_{\rm g} = E_{\rm g} \frac{Z_c-Z_v}{2Z_cZ_v} .
\end{equation}
Since $Z_c$ is typically smaller than $Z_v$, the QPA opens the band gap.
It is interesting that this term vanishes, if the Hermitian form \eqref{eq:Kotani} is replaced by the prescription of \myciteauthor{Casida1995},\cite{Casida1995} where the non-diagonal terms are evaluated at occupied energies only. 
Finally, it is worth mentioning that the result \eqref{eq:flat_band} is not affected by a degeneracy neither of the conduction nor the valence band. Both sides of the OEP equation would be multiplied by a common factor, which than cancels out. 

In summary, the QPA is a drastic approximation, which is generally unjustified except for states near the Fermi level. However, \textit{for the purpose of the OEP}, it can yield satisfactory results. This is manifest in the exact result for the HEG (Eq. \eqref{eq:Sham_HEG}), and the band gap correction (Eq. \eqref{eq:flat_band}) for our flat band model vanishes in the metallic limit. Further plausibility comes from the original derivation of Kotani \textit{et al.},\cite{Kotani2007} which is in spirit similar to that of the OEP equation. 
On the other hand, we can expect a failure of the QPA for both insulators and systems, where the exact treatment of the off-diagonal self-energy elements is important.
 
\section{\label{sec:Riemelmoser2021_3}	Computational details}

In Sec. \ref{sec:self-consistency cycle} we discuss how the self-consistency cycle is performed and describe details of our implementation in the PAW code \texttt{VASP} (Vienna \textit{Ab Initio} Simulation Package).\cite{Kresse1996} 
To motivate the need for an efficient self-consistency scheme, we remark that a single RPA-OEP calculation as typically performed in this work takes up roughly 100-3000 core hours, as compared to 10-60 core seconds for standard GGA calculations.

\subsection{Self-consistency cycle}\label{sec:self-consistency cycle}

As in any self-consistent scheme, a good starting point is important for fast convergence. The key trick is that a DFT potential (and the OEP is just that) usually converges faster with respect to the $k$-point mesh than other properties such as KS excitation energies. Hence, a good starting point for the OEP can by obtained from \textit{$v_{\rm xc}$ on a sparser $k$-point mesh.} Small differences on a denser $k$-point mesh can than be added with few further iterations.

\begin{figure}[!!tb]
\centering
\includegraphics [width=\linewidth,keepaspectratio=true] {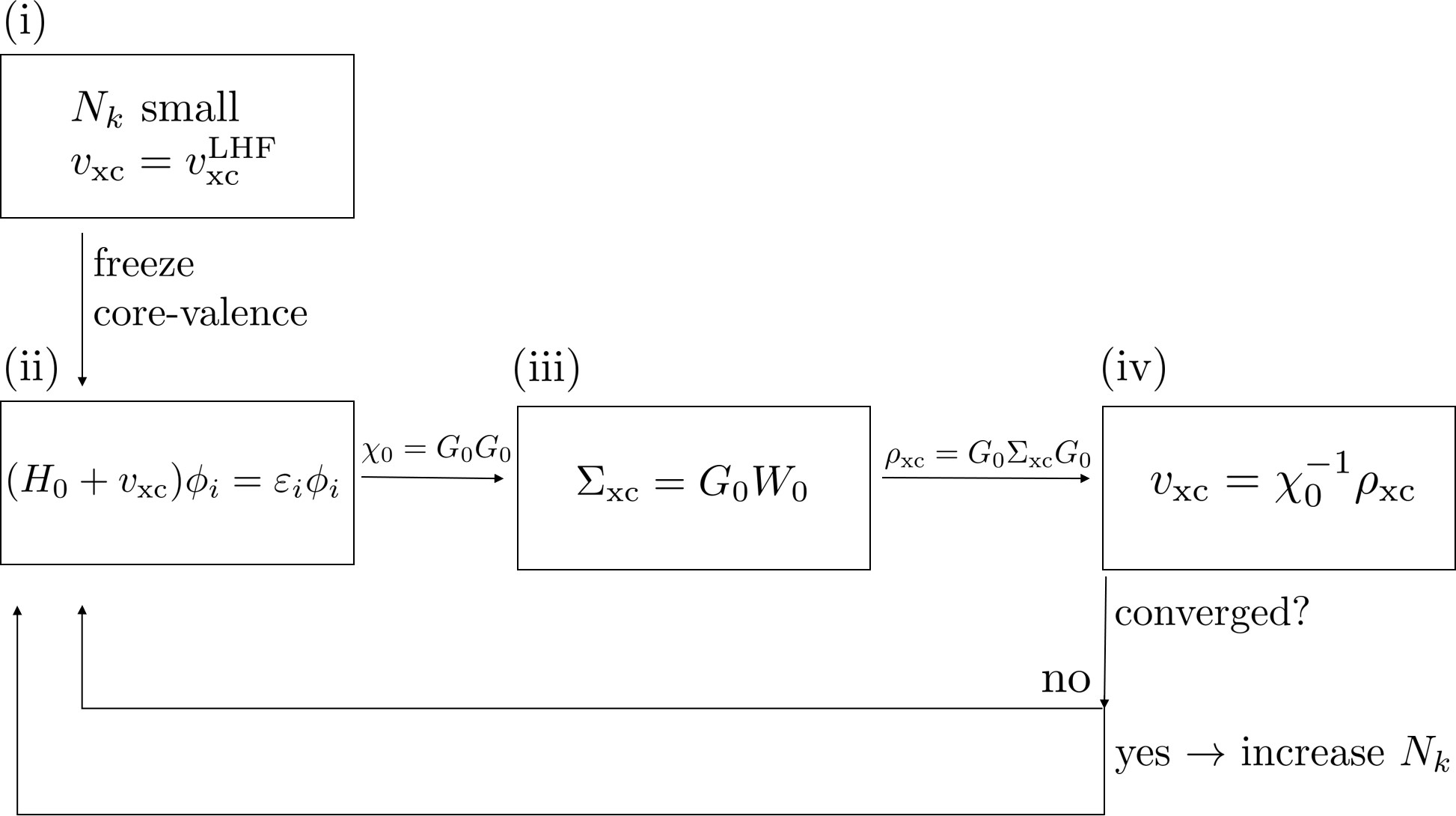}
\caption{Schematic representation of the self-consistency scheme for RPA-OEP calculations.}
\label{fig:self_consistency_scheme}
\end{figure}

The basic scheme is shown in Fig. \ref{fig:self_consistency_scheme}.  Starting with a sparse $k$-point grid, (i) we first perform a local Hartree-Fock\cite{DellaSala2001} calculation (LHF, an approximation to EXX-OEP) to obtain a starting point for the RPA-OEP. Following \myciteauthor{Klimes2014},\cite{Klimes2014} LHF yields an approximation to the core-valence interaction that is kept frozen during the rest of the calculation. The LHF calculation is performed in a one-shot way on-top of DFT reference orbitals.  Here, we use the GGA functional of Perdew, Burke and Ernzerhof\cite{Perdew1996,*Perdew1997}(PBE). The subsequent OEP self-consistency cycle is detailed in steps (ii) - (iv). The KS Hamiltonian using the RPA-OEP (in the first step the LHF) exchange-correlation potential is diagonalized, yielding $G_0$, from which $\chi_0$ and thereafter $\Sigma_{\rm xc}$ can be calculated. These steps parallel those of a routine $G_0W_0$ calculation, as detailed in Ref. \onlinecite{Liu2016}. Then, $\rho_{\rm xc}$ is evaluated in the natural orbital basis. The RPA-OEP potential is updated by letting $\chi_0^{-1}$ act on both sides of Eq. \eqref{eq:RPA_OEP}. In Appendix \ref{App:Riemelmoser2021_B}, more details on our implementation of step (iv) are given. The OEP is inserted back in the KS equation (ii), and the loop (ii)-(iv) is repeated until some convergence criterion is met. 
Then, we switch to the denser $k$-point mesh and re-enter the self-consistency loop. 

For QPA-RPA-OEP we use the implementation of Eq. \eqref{eq:QPA_RPA_OEP}	as presented in Ref. \onlinecite{Klimes2014}. It slightly differs from the current implementation in the way that frequency integrations are performed in step (iii) and uses the Kohn-Sham rather than the natural orbital basis in step (iv). To enable a fair comparison, we have carefully converged both calculations with respect to the relevant parameters (number of frequency points and number of unoccupied bands) to 10 meV accuracy in the OEP band gaps. 

Turning now to the performance of our self-consistency scheme, we find that switching from a $\Gamma$-centered $4 \times 4 \times 4$ $k$-point 
mesh to a $\Gamma$-centered $6\times 6\times 6$ $k$-point mesh, a single iteration is usually enough to converge the OEP band gaps within 10 meV. As a safety precaution, however, we always perform two iterations. The only approximation made is that the core-valence interaction terms are interpolated to the denser $k$-point mesh and that core-electrons are kept frozen at the level of the PBE reference. 

\begin{figure}[!!tb]
\centering
\includegraphics [angle=-90,width=\linewidth,keepaspectratio=true] {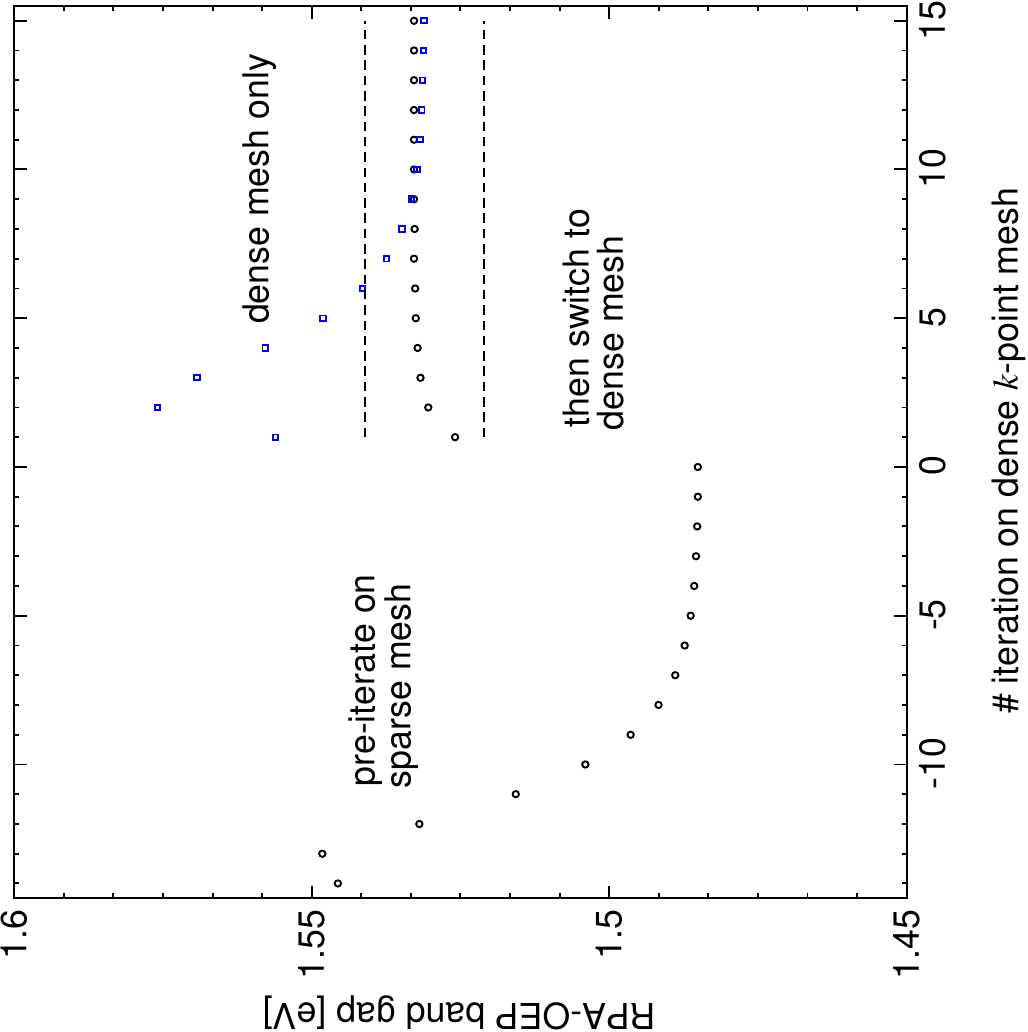}
\caption{Convergence of the RPA-OEP band gap for SiC, with respect to the number of iterations in the self-consistency cycle. Squares: 15 iterations on a $6 \times 6 \times 6$ $k$-point mesh, starting from LHF. Circles: 15 iterations on a $4 \times 4\times 4$ $k$-point mesh, then 15 further iterations on a $6 \times 6 \times 6$  $k$-point mesh. Dashed lines indicate a tolerance region of $\pm$10 meV.}
\label{fig:self_consistency_convergence}
\end{figure}

Fig. \ref{fig:self_consistency_convergence} demonstrates this for the exemplary case of SiC. Clearly, the stated approximation is excellent, as both procedures converge towards nearly the same value (here within 2 meV, generally we found $\lesssim$ 10 meV accuracy, compare also Table \ref{tab:k-points}). Furthermore, it can be seen that the pre-iteration procedure speeds up the self-consistency cycle on the more expensive $6\times 6\times 6$ $k$-point mesh, as a single iteration is enough to converge the band gap to 10 meV accuracy versus 6 iterations if starting directly from LHF.  

\begin{table}[!tb] 
\begin{ruledtabular}
\caption{RPA-OEP excitation energies for SiC from the valence band maximum at the $\Gamma$-point to the conduction band minimum at the indicated $k$-point (in eV). The first column shows results using a $4\times 4\times 4$ $\Gamma$-centered $k$-point mesh. The second column shows results for a $6 \times 6 \times 6$ mesh, and the third and fourth columns correspond to our interpolation method after 2 and 15 iterations on the dense mesh respectively, compare Fig. \ref{fig:self_consistency_convergence}. The band gap as reported in the following tables is marked bold. }
\label{tab:k-points}
\begin{tabular}{lrrrr}
 &  $4 \times 4 \times 4$ & $6 \times 6 \times 6$ &\multicolumn{2}{c}{interpolation method}   \\
  \cline{4-5}\\\\[-3.\medskipamount]  
  & & & 2 iterations & 15 iterations \\
\hline \\\\[-3.\medskipamount]
$\Gamma$ & 6.501  & 6.505 & 6.512  & 6.515\\
L & 5.606  & 5.628 & 5.633  & 5.635\\
X & 1.485  & 1.531 & \textbf{1.530}  & 1.533 
\end{tabular}
\end{ruledtabular}
\end{table}

\section{\label{sec:Riemelmoser2021_4} Applications}
\subsection{Applied settings}

In the present work, we consider 15 semiconductors and insulators listed in Table \ref{tab:settings}. Calculations are performed at the experimental lattice constant, where for materials that crystallize in the wurtzite structure under normal conditions\cite{Madelung2004} (GaN, ZnO, CdS), the zinc-blende structure was chosen to allow for comparison with previous calculations. 

RPA-OEP calculations are done using the \texttt{VASP} code as described previously in Sec. \ref{sec:Riemelmoser2021_3}. The plane wave cutoffs for the orbital basis $E_{\rm max}$ (\texttt{ENCUT} in \texttt{VASP}) are listed in Table \ref{tab:settings}. For the response function, we use a smaller cutoff $E_{\rm max}^{\chi}= E_{\rm max}/2$ (\texttt{ENCUTGW} in \texttt{VASP}). This is possible, as the convergence of the OEP tends to be better behaved with respect to $E_{\rm max}^{\chi}$ than single $GW$ self-energies, which often require even an extrapolation towards infinite basis, see e.g. Ref. \onlinecite{Klimes2014a}. Better convergence for OEP is expected, since the OEP equation \eqref{eq:RPA_OEP} takes essentially a self-energy average (the basis set incompleteness error also falls off as ${E^{\chi}_{\rm max}}^{-3/2}$, but with a different prefactor\cite{Klimes2014}). Nevertheless, the finite energy cutoff is the dominant error in our calculations, typically underestimating the band gaps by up to 50 meV. The two RPA-OEP methods also differ slightly in details of the implementation, see Sec. \ref{sec:Riemelmoser2021_3}. All in all, we estimate that the accuracy of the RPA-OEP band gaps is better than 100 meV. Nevertheless, we list them to higher precision, as the effect of the QPA, i.e. the differences $\Delta E_{\rm g}$, are estimated to be more accurate, within 50 meV.

EXX-OEP calculations are performed using the shift algorithm of \myciteauthor{Kuemmel2003}\cite{Kuemmel2003,Kuemmel2003a} as presented in Ref. \onlinecite{Klimes2014}. As unoccupied states are not required here, the EXX-OEP band gaps are more accurate, with errors well below 50 meV, and the dominant error is now related to the finite $k$-point mesh.

\begin{table}[!tb] 
\begin{ruledtabular}
\caption{Chosen settings for the OEP calculations. Strukturbericht designations of the chosen crystal structures are listed in parenthesis. Short-hands \_sv for the PAW potentials indicate treatment of the outermost $s$ and $p$ core states as valence. Moreover, short-hands \_h and \_nc indicate a very accurate treatment of high energy states. Orbital plane wave cutoffs $E_{\rm max}$ are given in eV. Values for the experimental lattice constant $a_0$ (in \AA) are taken from Ref. \onlinecite{Grueneis2014}, unless stated otherwise.}
\label{tab:settings}
\begin{tabular}{lcccl}
 &  & PAW potentials & $E_{\rm max}$ & $a_0$ \\ 
\hline \\\\[-3.\medskipamount]
C & (A4)  & C\_h\_GW & 742  & 3.567 \\ 
Si & (A4) & Si\_sv\_GW & 548  & 5.431 \\
SiC & (B3)  & Si\_sv\_GW C\_h\_GW & 742  & 4.358 \\ 
BN & (B3) & B\_h N\_h\_GW & 756 &  3.616\\
AlP & (B3) & Al\_sv\_GW P\_sv\_GW & 553  & 5.463 \\
AlAs & (B3) & Al\_sv\_GW As\_sv\_GW\_nc & 674  & 5.661 \\
AlSb & (B3) & Al\_sv\_GW Sb\_sv\_GW\_nc & 607  & 6.136 \\
GaN & (B3) & Ga\_sv\_GW\_nc N\_h\_GW & 802  & 4.531 \cite{Madelung2004} \\
GaP & (B3) & Ga\_sv\_GW\_nc P\_sv\_GW & 802  & 5.451 \\
InP & (B3) & In\_sv\_GW\_nc P\_sv\_GW & 604  & 5.869 \\ 
InSb & (B3) & In\_sv\_GW\_nc Sb\_sv\_GW\_nc & 607  & 6.479 \\ 
ZnO & (B3) & Zn\_sv\_GW\_nc O\_h\_GW & 802  & 4.580 \cite{Liu2016}\\ 
ZnS & (B3) & Zn\_sv\_GW\_nc S\_GW\_nc & 802  & 5.409 \\
CdS & (B3) & Cd\_sv\_GW\_nc S\_GW\_nc & 658  & 5.818 \\
MgO & (B1) & Mg\_sv\_GW\_nc O\_h\_GW & 822  & 4.211  \\  
\end{tabular}
\end{ruledtabular}
\end{table}

\subsection{OEP band gaps}\label{sec:OEP band gaps}

In Table \ref{tab:band_gaps}, we report OEP excitation energies from the valence band maximum (VBM) at the $\Gamma$-point to the conduction band minimum (CBM) at the $\Gamma$, L, and X-point, respectively. For comparison, we also include Kohn-Sham excitation energies obtained using the PBE functional.

\begin{table}[!tb] 
\begin{ruledtabular}
\caption{Kohn-Sham excitation energies obtained with the PBE functional, EXX-OEP as well as RPA-OEP with and without the quasiparticle approximation (QPA, exact). The energy excitations between the VBM at the $\Gamma$-point and the CBM at the indicated $k$-point are given in eV. All calculations are performed self-consistently. Wherever InSb is predicted to be metallic, we report a negative band gap (see text). }
\label{tab:band_gaps}
\begin{tabular}{llrrrr}
 & & PBE & EXX & \multicolumn{2}{c}{RPA} \\
  \cline{5-6}\\\\[-3.\medskipamount]
 &  &  &  & QPA & exact \\ %
\hline \\\\[-3.\medskipamount]
C     & $\Gamma$  &  5.60  &	 6.21&6.04	 &5.75  \\
      & L         &  8.47  &     9.08&8.89	 &8.62  \\
      & X         &  4.76  &     5.42&5.00	 &4.89  \\
      &            &         &                         \\[-1.\medskipamount]
Si    & $\Gamma$  &  2.55  &     3.13&2.69	 &2.68  \\
      & L         &  1.53  &     2.32&1.69	 &1.64  \\ 
      & X         &  0.70  &     1.34&0.74	 &0.74  \\
      &            &         &                         \\[-1.\medskipamount]
SiC   & $\Gamma$  &  6.21  &     7.49&6.75	 &6.51  \\
      & L         &  5.41  &     6.38&5.81	 &5.63  \\
      & X         &  1.34  &     2.37&1.60	 &1.53  \\
      &            &         &                         \\[-1.\medskipamount]
BN    & $\Gamma$  &  8.82  &     9.80&9.49	 &9.12  \\
      & L         & 10.13  &    11.19&10.69	 &10.37 \\
      & X         &  4.46  &     5.57&4.92	 &4.70  \\
      &            &         &                         \\[-1.\medskipamount]
AlP   & $\Gamma$  &  3.27  &     4.35&3.60	 &3.50  \\
      & L         &  2.78  &     3.51&2.97	 &2.92  \\
      & X         &  1.57  &     2.26&1.70	 &1.71  \\
      &            &         &                         \\[-1.\medskipamount]
AlAs  & $\Gamma$  &  2.02  &     3.10&2.32	 &2.28  \\
      & L         &  2.13  &     2.86&2.32	 &2.31  \\
      & X         &  1.44  &     2.11&1.61	 &1.62  \\
      &            &         &                         \\[-1.\medskipamount]
AlSb  & $\Gamma$  &  1.59  &     2.49&1.73	 &1.73  \\
      & L         &  1.34  &     1.94&1.43	 &1.45  \\
      & X         &  1.22  &     1.73&1.29	 &1.34  \\
      &            &         &                         \\[-1.\medskipamount]
GaN   & $\Gamma$  &  1.61  &     3.10&2.10	 &1.84  \\
      & L         &  4.52  &     5.81&4.93	 &4.67  \\
      & X         &  3.34  &     4.65&3.68	 &3.51  \\
      &            &         &                         \\[-1.\medskipamount]
GaP   & $\Gamma$  &  1.81  &     2.97&2.12	 &1.99  \\
      & L         &  1.66  &     2.45&1.82	 &1.76  \\
      & X         &  1.65  &     2.15&1.67	 &1.67  \\
      &            &         &                         \\[-1.\medskipamount]
InP   & $\Gamma$  &  0.69  &     1.68&0.84	 &0.79  \\
      & L         &  1.49  &     2.19&1.55	 &1.53  \\
      & X         &  1.77  &     2.26&1.76	 &1.77  \\
      &            &         &                         \\[-1.\medskipamount]
InSb  & $\Gamma$  &  -0.19  &    0.71&-0.16	 &-0.16 \\
      & L         &  0.54  &     1.11&0.53	 & 0.54 \\
      & X         &  1.34  &     1.67&1.28	 & 1.30 \\
	&            &         &                        &\\[-1.\medskipamount]
ZnO   & $\Gamma$  &  0.68  &     2.90&1.46	 &1.20  \\
      & L         &  5.46  &     7.43&6.18	 &5.91  \\
      & X         &  5.27  &     7.52&5.99	 &5.77  \\
      &            &         &                         \\[-1.\medskipamount]
ZnS   & $\Gamma$  &  2.09  &     3.35&2.36	 &2.25  \\
      & L         &  3.30  &     4.34&3.53	 &3.46  \\
      & X         &  3.41  &     4.24&3.50 	 &3.49  \\
	&            &         &                        \\[-1.\medskipamount]
CdS   & $\Gamma$  &  1.15  &     2.20&1.20  	 &1.17  \\
      & L         &  3.02  &     3.87&3.04  	 &3.04  \\
      & X         &  3.54  &     4.30&3.56  	 &3.56  \\
      &            &         &                         \\[-1.\medskipamount]
MgO   & $\Gamma$  &  4.74  &     6.55&5.58  	 &5.23  \\
      & L         &  7.91  &     9.71&8.66  	 &8.36  \\
      & X         &  9.12  &    10.88&9.74  	 &9.50  \\
\end{tabular}
\end{ruledtabular}
\end{table}
We first note that our numerical results for EXX-OEP and QPA-RPA-OEP excitation energies are generally in good agreement with those of \myciteauthor{Klimes2014}.\cite{Klimes2014} Slight differences for SiC, AlP, and AlAs are due to our use of more accurate pseudo-potentials for Si, P and As, respectively. 

In the following, we discuss first general trends and thereafter quantitative results and specific materials. Generally, we observe the following order: PBE yields the smallest gaps, and EXX-OEP the largest. When screening in form of the QPA-RPA is taken into account, the gaps close substantially with respect to EXX-OEP. When the QPA is lifted, taking dynamical screening properly into account, the band gaps are further closed towards PBE. It is reassuring that similarly, the static COHSEX approximation tends to overestimate band gaps.\cite{Hybertsen1986}

In summary, we obtain the following order for Kohn-Sham band gaps
\begin{equation*}
\text{PBE} \leq \text{RPA-OEP} \leq \text{QPA-RPA-OEP} \ll \text{EXX-OEP} .  
\end{equation*}
Moreover, we find that the effect of the QPA is stronger at the $\Gamma$-point than at the L-point and weakest at the X-point. This brings the dispersion of the RPA-OEP conduction bands closer to PBE.

Turning now to the discussion of specific systems,  there are two groups of materials, where exact RPA-OEP differs most from QPA-RPA-OEP. (i) Large effects are seen for all insulators, where the indirect $\Gamma\rightarrow X$ gaps are closed by 0.1 eV (C), 0.2 eV (BN) and 0.35 eV (MgO) respectively. (ii) The band gaps are also much reduced in those materials, where the accurate treatment of the off-diagonal elements of $\Sigma_{\rm xc}$ is important.\cite{Kotani2007,Shishkin2007a} Here, the direct $\Gamma \rightarrow \Gamma$ gaps are closed by 0.25 eV for both ZnO and GaN. The off-diagonal elements also play a large role for MgO,\cite{Shishkin2007a} which further explains why the effect is even stronger here than for the other insulators.

Generally, it can be observed that group (ii) is characterized by a large electronegativity difference of the constituents: compare the effects of the QPA for ZnO/ZnS, GaN/GaP, the series C-BN-MgO, and note also that the QPA has small effects for the almost electroneutral systems AlP, AlAs and AlSb. Broadly speaking, the character of the VBM and the CBM is rather dissimilar in ionic systems. This is reflected in the difference between the $Z$-factors of the CBM and the VBM, which appears also in our model estimate for $\Delta_{\text g}$, see Eq. \eqref{eq:flat_band}. Similarly, it has been shown that for ionic systems, range-separated hybrid functionals can struggle to reproduce optical properties as calculated by more accurate MBPT methods,\cite{Chen2012,*Chen2013} even if the parameters of the hybrids are optimally tuned.\cite{Tal2020} This underlines the importance of dynamical screening effects for ionic systems.

For AlSb, the QPA slightly closes the indirect band gap by 0.05 eV, in contradiction to the general trend. This effect is, however, within our estimated accuracy, and the dispersion is closer to PBE for exact RPA-OEP as is generally the case.

Finally, we want to comment briefly on the interesting case of InSb, which is predicted to be metallic by PBE as well as by both RPA-OEP schemes. Alongside the metallic phase, a band inversion occurs, i.e. the three-fold degenerate $p$ bands (usually forming the top of the valence band) are then above the $s$ band at the $\Gamma$-point. Per convention, the excitation energies are still reported with respect to the $p$ bands, yielding a ``negative band gap''.  We find that this negative band gap is virtually the same for PBE, RPA-OEP and QPA-RPA-OEP.

\subsection{Band structures and potentials}

\begin{figure}[!!tb]
\centering
\subfloat{\includegraphics [angle=-90,width=\linewidth,keepaspectratio=true] {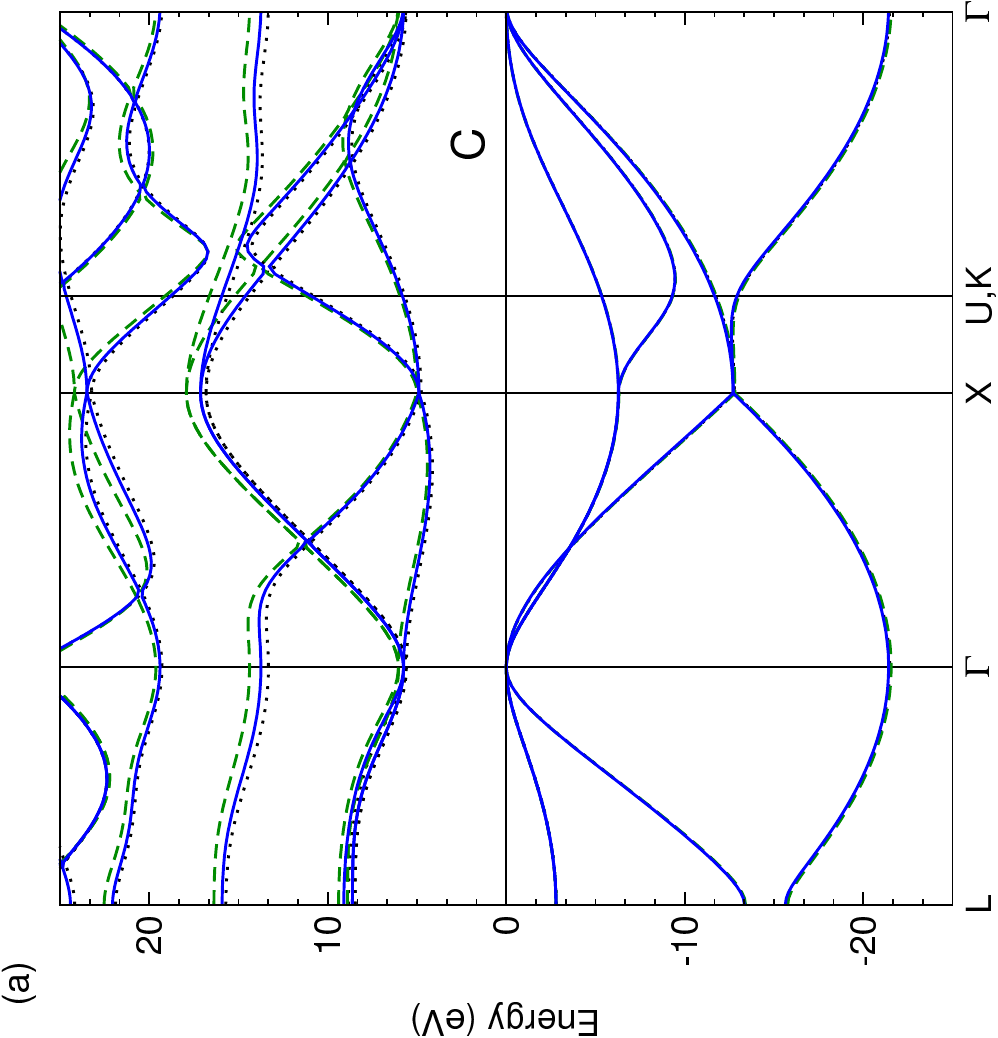}\label{fig:bands_a}}
\par\medskip
\subfloat{\includegraphics [angle=-90,width=\linewidth,keepaspectratio=true] {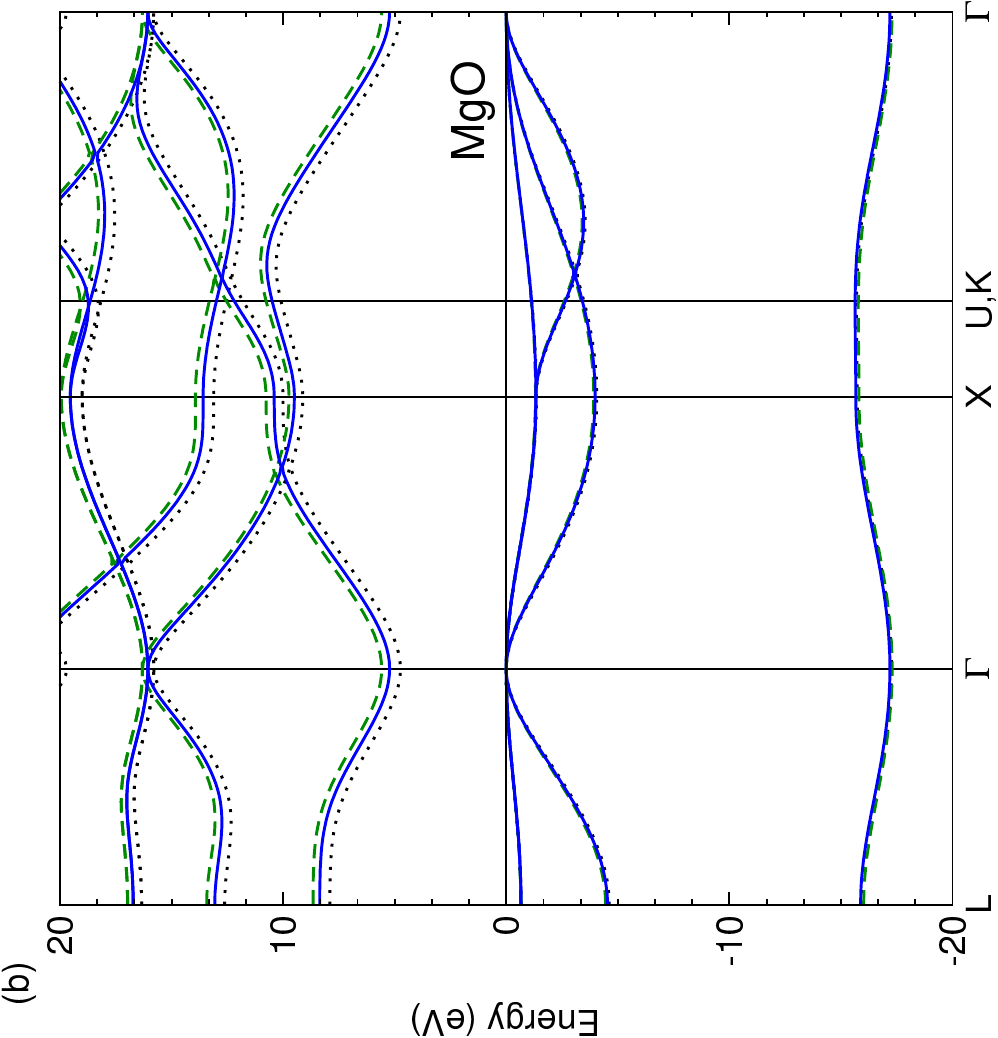}\label{fig:bands_b}}
\caption{Kohn-Sham band structures for (a) C and (b) MgO. Dotted (black) lines represent PBE, dashed (green) lines QPA-RPA-OEP and solid (blue) lines exact RPA-OEP, respectively. Valence band maxima at the $\Gamma$-point have been aligned at zero.}
\label{fig:bands}
\end{figure}

To further substantiate the results of Sec. \ref{sec:OEP band gaps}, we now discuss two representative systems, C and MgO, in greater detail. To this effect, we compare RPA-OEP band structures and KS potentials in real space to the respective PBE references. Starting with C, we find from the band structure shown in Fig. \ref{fig:bands}(a) that the valence bands of both RPA-OEP methods are very close to the PBE ones. The conduction bands from exact RPA-OEP resemble those from PBE, up to a rigid up-shift of about 150 meV, whereas the conduction bands from QPA-RPA-OEP exhibit broader dispersion and band width. Fig. \ref{fig:bands}(b) shows that this broader dispersion for QPA-RPA-OEP is less pronounced in MgO, where the band structures are close for all three methods, again up to rigid shifts of the conduction bands. Hence, it is difficult to draw general conclusions, except that for typical semiconductors and insulators as investigated in this work the PBE band structures are close to the exact RPA-OEP ones  (up to a rigid shift of the conduction bands), compare also Table \ref{tab:band_gaps}. This reaffirms the observations of Godby \textit{et al.},\cite{Godby1986,Godby1988} then made with respect to LDA rather than PBE.

\begin{figure}[!!tb]
\centering
\subfloat{\includegraphics [angle=-90,width=\linewidth,keepaspectratio=true] {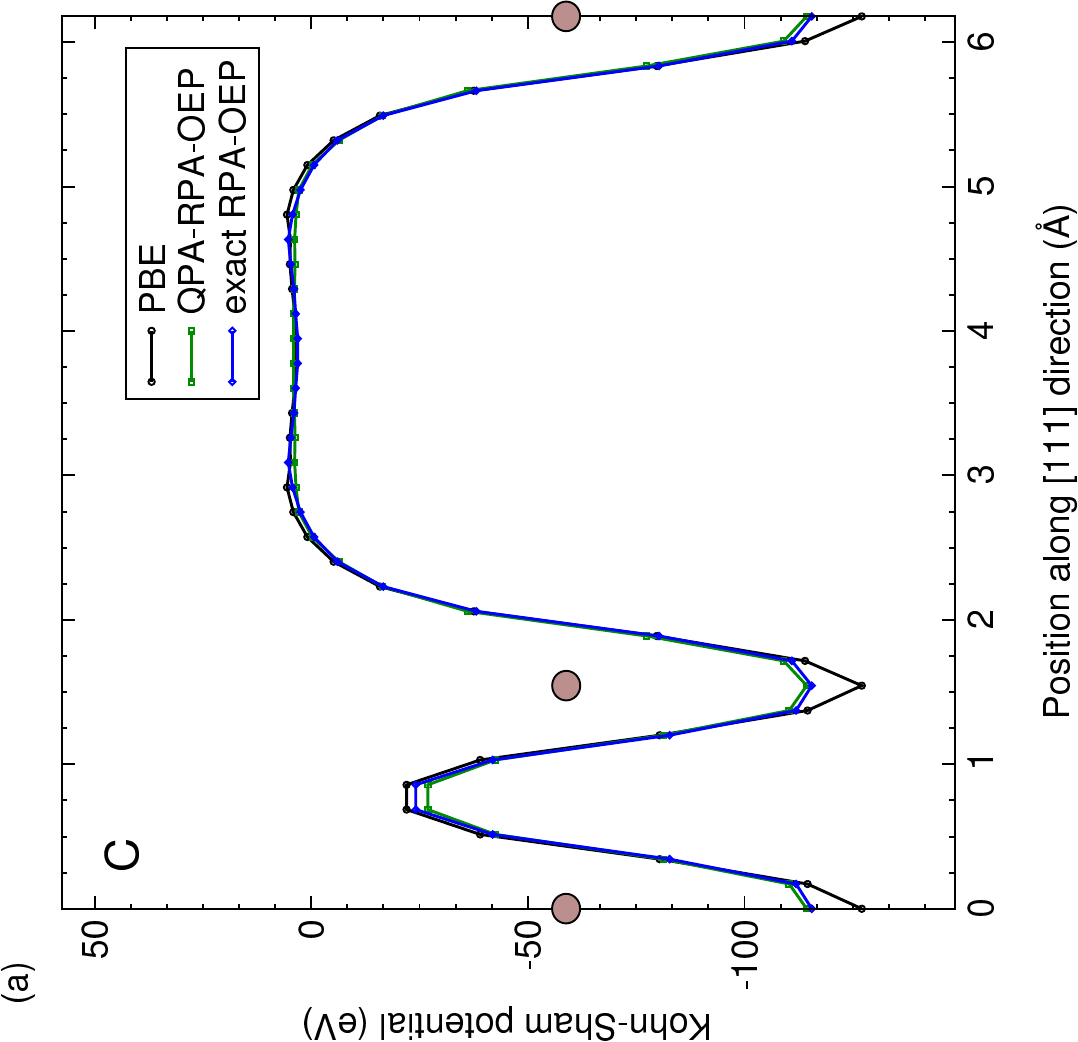}\label{fig:potentials_C}}
\par\medskip
\subfloat{\includegraphics [width=\linewidth,keepaspectratio=true] {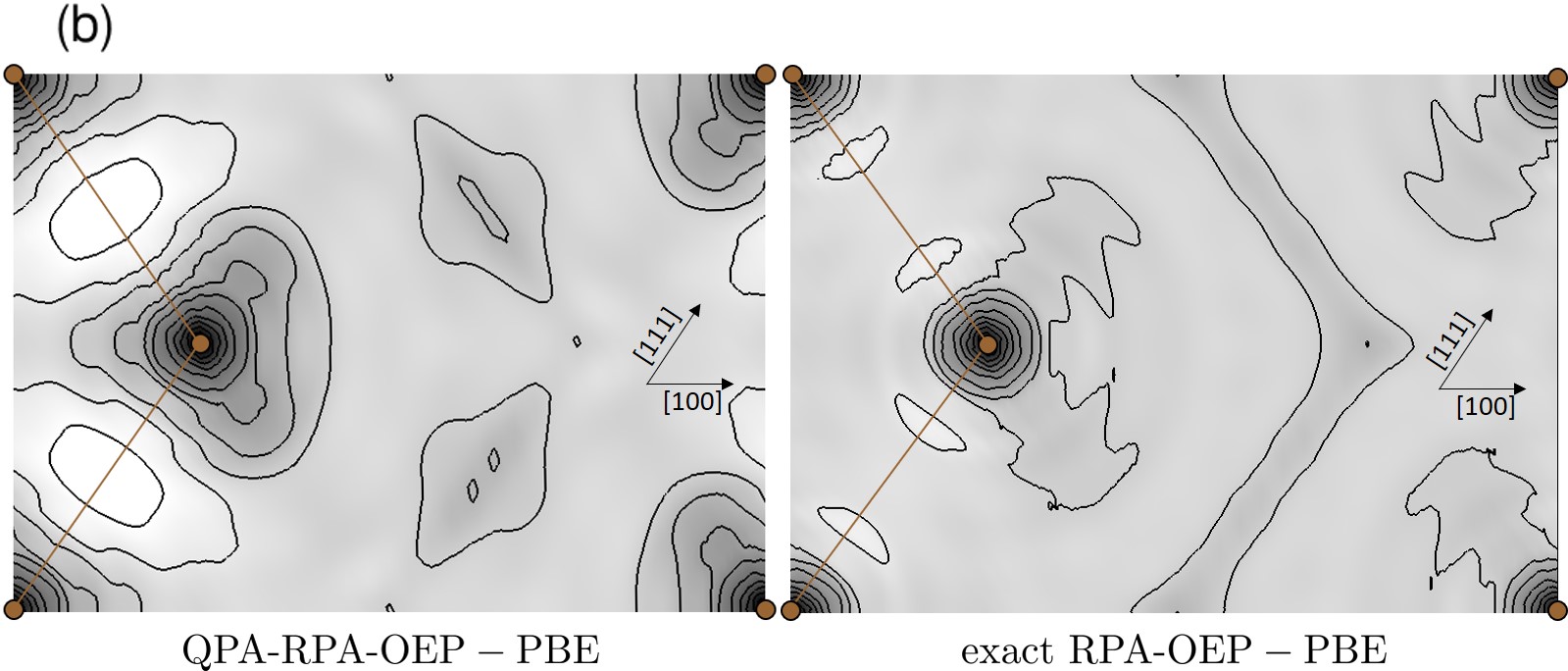}\label{fig:contour_C}}
\caption{Comparison of the KS potentials $v_{\rm KS}(\textbf{r})$ for C. The atom positions are indicated by small circles. (a) Total KS potentials along the [111] direction. Symbols indicate the finite real space grid, on which $v_{\rm KS}$ is evaluated.  (b) Contour plots for the differences $v_{\rm KS}^{\rm OEP}-v_{\rm KS}^{\rm PBE}$ in the $(0\bar{1}1)$ plane. Contour lines are drawn in steps of 1.5 eV, lighter areas indicate that $v_{\rm KS}^{\rm OEP}<v_{\rm KS}^{\rm PBE}$. Left: QPA-RPA-OEP, right: exact RPA-OEP. Straight lines indicate covalent bonds between carbon atoms. The potential differences have been interpolated between real space grid points using cubic splines.\cite{Momma2011}}
\label{fig:potentials_C}
\end{figure}

For the purpose of plotting the OEP potentials in real space, we redo the OEP calculations for C and MgO with an increased plane wave cutoff for the response function, $E_{\rm max}^{\chi}=2/3E_{\rm max}$. This has little effect on the band gaps, but helps to reduce numerical noise in the interstitial regions [for both C and MgO, the gaps open by $\lesssim 20 \text{ meV}$, the effect on the differences $\Delta E_{\rm g}$, see Eq. \eqref{eq:delta_g}, is negligible]. 
  
Fig. \ref{fig:potentials_C}(a) shows total KS potentials for C from the two RPA-OEP methods as well as from PBE along the [111] direction (atom positions are indicated by small circles). The KS potential from QPA-RPA-OEP agrees well with that reported by \myciteauthor{Klimes2014}.\cite{Klimes2014} Deeper KS potentials inside the PAW spheres in the current work are due to our use of a harder pseudo-potential. General features of the potentials are a well defined maximum at the center of the covalent bonds and fairly flat behavior in the interstitial regions. 

A contour plot of the differences  $v_{\rm KS}^{\rm OEP}-v_{\rm KS}^{\rm PBE}$ in the $(0\bar{1}1)$ plane is further shown in Fig. \ref{fig:potentials_C}(b). For C, similar comparisons have been made already by previous authors,\cite{Godby1988,Klimes2014} albeit with respect to LDA rather than PBE. This difference, however, does not affect the following qualitative analysis. As in Refs. \onlinecite{Godby1988,Klimes2014}, we observe that the RPA-OEP potentials are more attractive in the bonding regions, more so if dynamical screening is neglected in form of the QPA. Together with the fact that the potentials are fairly similar in the interstitial regions, this can explain the observed order for the band gaps.\cite{Godby1988} That is, the gap opens as the valence states, which are concentrated around the bonds, get lowered in energy relative to the conduction states.
Interestingly, Fig. \ref{fig:potentials_C}(b) shows that the KS potential from exact RPA-OEP has noticeable angular structure near the anti-bonding regions, when compared to the more spherically symmetric PBE potential. This structure is absent in QPA-RPA-OEP, compare also respective figures in Refs. \onlinecite{Godby1988,Klimes2014}. 

\begin{figure}[!!tb]
\centering
\subfloat{\includegraphics [angle=-90,width=\linewidth,keepaspectratio=true] {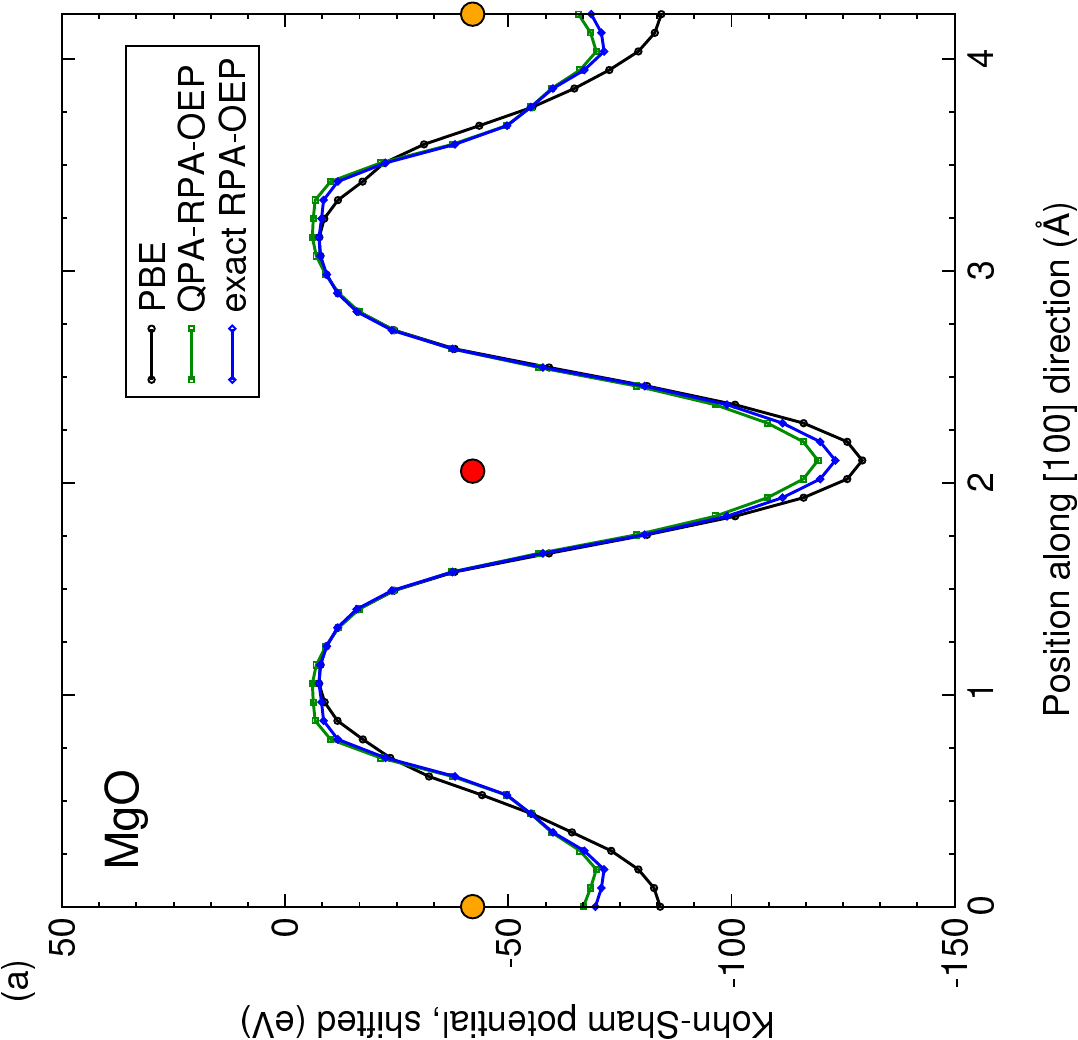}\label{fig:potentials_MgO}}
\par\medskip
\subfloat{\includegraphics [width=\linewidth,keepaspectratio=true] {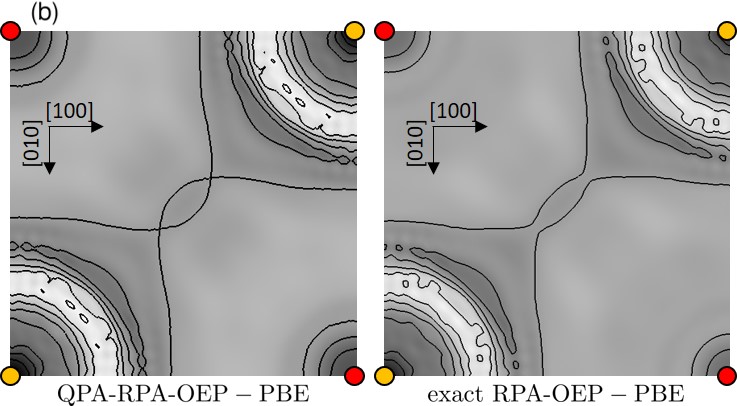}\label{fig:bands_b}}
\caption{Comparison of the KS potentials $v_{\rm KS}(\textbf{r})$ for MgO. The atom positions are indicated by light (orange) and dark (red) circles for Mg and O, respectively. For easier comparison, OEP potentials have been rigidly shifted to better fit PBE in the interstitial regions ($v_{\rm KS}^{\rm OEP}$ is only determined up to a constant). (a) Total KS potentials along the [100] direction. Symbols indicate the finite real space grid, on which $v_{\rm KS}$ is evaluated. (b) Contour plots for the differences $v_{\rm KS}^{\rm OEP}-v_{\rm KS}^{\rm PBE}$ in the $(001)$ plane. Contour lines are drawn in steps of 3 eV, lighter areas indicate that $v_{\rm KS}^{\rm OEP}<v_{\rm KS}^{\rm PBE}$. Left: QPA-RPA-OEP, right: exact RPA-OEP. The potential differences have been interpolated between real space grid points using cubic splines.\cite{Momma2011} }
\label{fig:potentials_MgO}
\end{figure}

Turning now to MgO, Fig. \ref{fig:potentials_MgO}(a) shows the total KS potentials along the [100] direction. The PBE potential is relatively structureless and resembles a broad barrier between atomic potential wells. Both RPA-OEP methods significantly deviate from PBE in the vicinity of Mg atoms (light (orange) circles), the effect is again stronger in QPA-RPA-OEP.  Contour plots of the potential differences in the (001) plane show distinct radial structures, see Fig. \ref{fig:potentials_MgO}(b). Weaker, but more important, these ``rings'' exist also in the vicinity of O atoms (dark (red) circles), and have larger amplitudes in the static QPA-RPA-OEP. The KS potential from EXX-OEP shows similar radial structure that is even stronger (not shown). One is tempted to invoke the interpretation from above once again: relatively compact valence states get pulled down by the ``white rings'', whereas more delocalized conduction states are less effected by the oscillatory potential difference. Finally, as observed for C, we find that the exact RPA-OEP potential is less spherically symmetric around the atoms. However, the angular structure is not as pronounced here, since the rock-salt crystal structure constraints the potentials more strongly to spherical symmetry.

\subsection{Dielectric constants from RPA-OEP}

\begin{table}[!tb] 
\begin{ruledtabular}
\caption{Ion-clamped, macroscopic dielectric constants in the independent particle approximation obtained from PBE as well as RPA-OEP with and without the quasiparticle approximation (QPA, exact). The calculations use a $14 \times 14 \times 14$ $\Gamma$-centered $k$-point mesh. Experimental dielectric constants are taken from Ref. \onlinecite{Shishkin2007a}, unless stated otherwise. Footnotes indicate that experiments were performed at wurtzite crystal structure rather than zinc-blende. Mean relative errors (MRE) and mean absolute relative errors (MARE) are reported as well.}
\label{tab:dielectric_IP}
\begin{tabular}{lrrrl}
& PBE & \multicolumn{2}{c}{RPA} & expt. \\
\cline{3-4}\\\\[-3.\medskipamount] 
& & QPA & exact &\\
\hline \\\\[-3.\medskipamount]
C    &6.0	 & 5.7	  & 5.8	 & \hspace{5pt}5.7 \\
Si   &13.5	 & 12.9	  & 13.1	 & 11.9 \\
SiC  &7.2	 & 6.8	  & 7.0	 & \hspace{5pt}6.5 \\
BN   &4.7	 & 4.5	  & 4.6	 & \hspace{5pt}4.5 \\
AlP  &8.8	 & 8.4	  & 8.4	 & \hspace{5pt}7.5 \\
AlAs &9.9	 & 9.3	  & 9.3	 & \hspace{5pt}8.2\cite{Madelung2004} \\
AlSb &12.2	 & 11.7	  & 11.6	 & 10.2\cite{Madelung2004} \\
GaN  &6.2	 & 5.6	  & 5.9	 & \hspace{5pt}$5.3^{a}$ \\
GaP  &10.9	 & 10.4	  & 10.6	 & \hspace{5pt}9.1\cite{Madelung2004} \\
InP  &11.3	 & 11.0	  & 11.1	 & 10.9\cite{Madelung2004} \\
ZnO  &5.4	 & 4.2	  & 4.5	 & \hspace{5pt}$3.7^{a}$ \\
ZnS  &6.3	 & 6.0	  & 6.1	 & \hspace{5pt}5.1 \\
CdS  &6.4	 & 6.4	  & 6.4	 & \hspace{5pt}$5.3^{a}$ \\
MgO  &3.2	 & 3.0	  & 3.1	 & \hspace{5pt}3.0 \\
\hline \\\\[-3.\medskipamount]
MRE & 16\% & 9\% & 11\% \\
MARE & 16\% & 9\% & 11\%\\
\hline \\\\[-3.\medskipamount]
\multicolumn{4}{l}{$^a$wurtzite structure} \\
\end{tabular}
\end{ruledtabular}
\end{table}
\begin{table}[!tb] 
\begin{ruledtabular}
\caption{Like Table \ref{tab:dielectric_IP}, but with dielectric constants in the RPA.}
\label{tab:dielectric_RPA}
\begin{tabular}{lrrrl}
& PBE & \multicolumn{2}{c}{RPA} & expt. \\
\cline{3-4}\\\\[-3.\medskipamount] 
& & QPA & exact &\\
\hline \\\\[-3.\medskipamount]
C    & 5.5 & 5.3	  &5.4   & \hspace{5pt}5.7 \\
Si   & 12.1& 11.6	  &11.7  & 11.9 \\
SiC  & 6.6 & 6.2	  &6.3   & \hspace{5pt}6.5 \\
BN   & 4.4 & 4.1	  &4.2   & \hspace{5pt}4.5 \\
AlP  & 7.6 & 7.2	  &7.3   & \hspace{5pt}7.5 \\
AlAs & 8.6 & 8.1	  &8.1   & \hspace{5pt}8.2\cite{Madelung2004} \\
AlSb & 10.7& 10.3	  &10.2  & 10.2\cite{Madelung2004} \\
GaN  & 5.7 & 5.2	  &5.5   & \hspace{5pt}$5.3^a$ \\
GaP  & 9.7 & 9.2	  &9.4   & \hspace{5pt}9.1\cite{Madelung2004} \\
InP  & 10.2& 9.8	  &9.9   & 10.9\cite{Madelung2004} \\
ZnO  & 5.1 & 3.9	  &4.2   & \hspace{5pt}$3.7^a$ \\
ZnS  & 5.6 & 5.3	  &5.4   & \hspace{5pt}5.1 \\
CdS  & 5.8 & 5.7	  &5.7   & \hspace{5pt}$5.3^a$ \\
MgO  & 3.0 & 2.8	  &2.9   & \hspace{5pt}3.0 \\
\hline \\\\[-3.\medskipamount]
MRE & 5\% & 0\% & 0\% \\
MARE & 7\% & 5\% & 5\%\\
\hline \\\\[-3.\medskipamount]
\multicolumn{4}{l}{$^a$wurtzite structure} \\
\end{tabular}
\end{ruledtabular}
\end{table}

\begin{table}[!tb] 
\begin{ruledtabular}
\caption{Like Table \ref{tab:dielectric_IP}, but with dielectric constants including vertex corrections via the LDA exchange-correlation kernel, see Eq. \eqref{eq:local_vertex}. }
\label{tab:dielectric_DFT}
\begin{tabular}{lrrrl}
& PBE & \multicolumn{2}{c}{RPA} & expt. \\
\cline{3-4}\\\\[-3.\medskipamount] 
& & QPA & exact &\\
\hline \\\\[-3.\medskipamount]
C    & 5.8  & 5.5   & 5.7  & \hspace{5pt}5.7 \\
Si   & 12.8 & 12.3  & 12.4 &            11.9 \\
SiC  & 6.9  & 6.5   & 6.7  & \hspace{5pt}6.5 \\
BN   & 4.6  & 4.3   & 4.4  & \hspace{5pt}4.5 \\
AlP  & 8.0  & 7.6   & 7.7  & \hspace{5pt}7.5 \\
AlAs & 9.2  & 8.6   & 8.6  & \hspace{5pt}8.2\cite{Madelung2004} \\
AlSb & 11.4 & 10.9  & 10.8 &            10.2\cite{Madelung2004} \\
GaN  & 5.9  & 5.4   & 5.7  &\hspace{5pt}$5.3^a$ \\
GaP  & 10.2 & 9.7   & 9.9  & \hspace{5pt}9.1\cite{Madelung2004} \\
InP  & 10.7 & 10.3  & 10.4 &            10.9\cite{Madelung2004} \\
ZnO  & 5.3  & 4.1   & 4.4  &\hspace{5pt}$3.7^a$ \\
ZnS  & 5.8  & 5.5   & 5.6  & \hspace{5pt}5.1 \\
CdS  & 6.0  & 6.0   & 6.0  &\hspace{5pt}$5.3^a$ \\
MgO  & 3.2  & 2.9   & 3.0  & \hspace{5pt}3.0 \\
\hline \\\\[-3.\medskipamount]
MRE & 10\% & 3\% & 5\% \\
MARE & 11\% & 5\% & 6\%\\
\hline \\\\[-3.\medskipamount]
\multicolumn{4}{l}{$^a$wurtzite structure} \\
\end{tabular}
\end{ruledtabular}
\end{table}

In this section, we investigate the effect of the QPA on dielectric constants. The potentials from both RPA-OEP methods are used to calculate ion-clamped, macroscopic dielectric constants $\epsilon_{\infty}$, 
\begin{equation}
\epsilon_{\infty}^{-1} = \lim_{\omega\to 0,\textbf{q}\to 0} \epsilon^{-1}(\textbf{q},\textbf{q},	\omega) .
\end{equation}
The dielectric constants are evaluated (i) from the independent particle response function $\chi_0$, (ii) from the RPA response function, compare Eq. \eqref{eq:RPA_screening},
and (iii) including vertex corrections in the form of local field effects. For the latter, one has to solve the Dyson-like equation 
\begin{equation}\label{eq:local_vertex}
\chi = \chi_0 + \chi_0(V+f_{\rm xc})\chi .
\end{equation}
The exchange-correlation kernel $f_{\rm xc}=\delta v_{\rm xc}/\delta n$ is approximated by its LDA version, for details of the calculations we refer to Ref. \onlinecite{Gajdos2006}. The results are collected in Tables \ref{tab:dielectric_IP}-\ref{tab:dielectric_DFT}, where the metallic InSb is excluded. For reference, we report also dielectric constants obtained from PBE. Where data for $\epsilon_{\infty}^{\rm RPA}$ is available, our dielectric constants from PBE agree well with those of Ref. \onlinecite{Shishkin2007}. Results for $\epsilon_{\infty}^{\rm RPA}$ from QPA-RPA-OEP are somewhat larger than those of Ref. \onlinecite{Klimes2014}. This discrepancy, which is largest for Si (10.9 in Ref. \onlinecite{Klimes2014} vs. our result of 11.6), can be traced back to better $k$-point convergence in the present work. 

In the following, we once again comment first on general trends before discussing specific materials and comparing to experiment. Generally, we find that dielectric constants from PBE are the largest, followed by exact RPA-OEP and QPA-RPA-OEP. This is not surprising, as the dielectric constants are roughly inverse proportional to the band gap. Comparing results from the RPA-OEP methods, we find that the effect of the QPA is reduced compared to the band gaps. That is, the relative differences $\Delta \epsilon_{\infty}/\epsilon_{\infty}$ are somewhat smaller than the corresponding relative differences for the band gap, $\Delta E_{\rm g}/E_{\rm g}$. Similarly, it was reported in Ref. \onlinecite{Shishkin2007} that the discrepancy between different $GW$-methods is reduced for the dielectric constants. Comparing the different kernels for a given KS potential, we find that the dielectric constants in the independent particle approximation are generally the largest, those in RPA the smallest, and the vertex corrected ones in-between.

Comparing now to experiment, we find that the dielectric constants in the independent particle approximation are generally larger than the experimental ones. If screening in the RPA is taken into account, the dielectric constants from PBE are much reduced, but still larger than experiment. Agreement with experiment is now quite good for both RPA-OEP methods (MARE 5\% each, no systematic errors can be discerned). Larger effects of the QPA are observed only for ZnO and GaN (the dielectric constant of MgO is quite small). If the LDA kernel is included, the dielectric constants are on average once again larger than the experimental ones, especially for PBE.

\section{\label{sec:Riemelmoser2021_5}Discussion}

In Sec. \ref{sec:Riemelmoser2021_4} we have presented results for band gaps and dielectric constants from both PBE and two RPA-OEP methods. In the following, we qualitatively discuss the approximations made and speculate on what exact Kohn-Sham theory would yield. Generally, the band gaps give more direct insight, whereas the interpretation of the dielectric constants is more difficult, as one has to disentangle approximations for the KS potential and the exchange-correlation kernel.

\subsection{Band gaps}

We first note that semi-local functionals like PBE have spurious self-interaction that delocalizes the orbitals and thus the exchange-correlation hole. Therefore, the eigenvalues are up-shifted compared to those of the exact KS potential. The effect is stronger for the more localized occupied orbitals, and it follows that the minimal gaps of (semi-)local functionals are generally too small. In fact, one can observe this trend quite consistently in small molecules, where an exact KS potential can be obtained as reference.\cite{Baerends2013} Unsurprisingly, the gaps are underestimated less by GGA than by LDA, as the self-interaction is even more severe for LDA.

RPA-OEP, on the other hand, differs from the true KS potential by neglecting diagrams corresponding to vertex corrections. The most important vertex corrections are (i) screened exchange diagrams (corrections to $\Sigma_{\rm xc}$) and (ii) ladder diagrams (corrections to $W$). The screened exchange diagrams cancel the self-interaction error inherent to the RPA, thus including them opens the band gap. The ladder diagrams provide additional screening, which closes the band gap. 

Previous work has shown that the ladder diagrams, albeit of higher order in the Coulomb interaction, are the most important corrections for ionization energies of both solids\cite{Grueneis2014} and molecules.\cite{Maggio2017} Moreover, \myciteauthor{Kutepov2016}\cite{Kutepov2016,Kutepov2017} has found that the band gaps of solids are reduced by the inclusion of vertex corrections in self-consistent $GW$ calculations. However, the picture is more complicated here as self-consistency introduces additional diagrams. Generally, one has to be careful in the comparison of $GW$ and OEP,\cite{Caruso2013} see also Sec. \ref{sec:model_QPA}. More direct evidence is given by the results of \myciteauthor{Hellgren2007},\cite{Hellgren2007} who have shown that the RPA-OEP excitation energies of atoms tend to be slightly larger than those of the exact KS potential.
In any case, the exact balance of screened exchange diagrams and ladder diagrams is expected to be strongly system dependent, though.\cite{Maggio2017,Grueneis2014}

Nevertheless, we believe that for the solids considered in the present work \textit{a good estimate for the true KS gap is probably in-between PBE and RPA-OEP}. This is satisfying insofar, as we can conclude that the exact RPA-OEP is thus closer to the true KS potential than QPA-RPA-OEP. We reiterate that also from a physical stand point, the exact RPA-OEP has to be considered the better theory, see Sec. \ref{sec:model_QPA}.
It is safer to say that for materials where PBE and both RPA-OEP methods are close, e.g. for Si, we can further conclude that all three methods yield decent approximations to the true KS gap. Moreover, we conjecture that the exact KS potential will also predict InSb to be metallic.  Naturally, the fact that experiments yield a small, but finite fundamental band gap of 0.2 eV for InSb\cite{Madelung2004} does not indicate a failure of the KS formalism. Rather, this qualitative discrepancy underlines the importance of the derivative discontinuity, which has to be taken into account to enable comparison with experiment. 

\begin{figure}[!!tb]
\centering
\includegraphics [width=\linewidth,keepaspectratio=true] {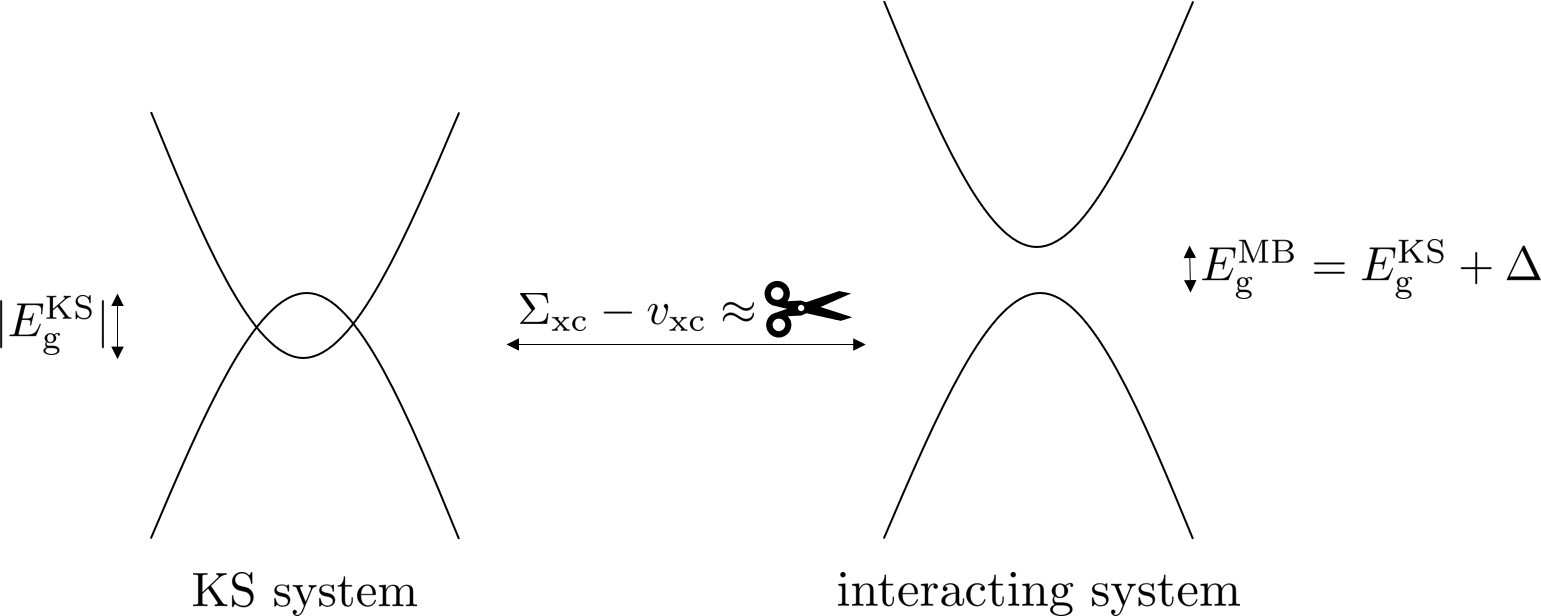}
\caption{Comic illustrating the band inversion caused by the derivative discontinuity $\Delta = \braket{\Sigma_{\rm xc} - v_{\rm xc}}$. In the ``scissors operator'' approximation,\cite{Godby1988} $\Delta$ simply rigidly shifts the conduction bands. If $\Delta$ is larger than the fundamental gap $E_{\rm g}^{\rm MB}$,  the KS band gap $E_{\rm g}^{\rm KS}$ is negative and a band inversion occurs alongside a metal-semiconductor phase transition.}
\label{fig:scissors}
\end{figure}

Finally, it is intuitively clear that a band inversion will occur alongside a negative band gap, if the derivative discontinuity can be approximated as a ``scissors operator'',\cite{Godby1988} see Fig. \ref{fig:scissors}. We have observed that the band inversion for InSb is incipient even in the first step of the self-consistency cycle. That is, the RPA natural orbitals (on-top of LHF) corresponding to the CBM show an occupancy much larger than zero, whereas those corresponding to the VBM are depleted (not shown). This is related to density fluctuations from the occupied Kohn-Sham orbitals into the unoccupied Kohn-Sham states. To achieve the corresponding charge density using a mean field method, the OEP potential needs to bring the conduction band down, eventually below the valence band. It is plausible that the band inversion will prevail even for the exact Kohn-Sham potential, given that these density fluctuations are described well enough by the RPA.

\subsection{Dielectric constants}

Let us assume for the moment that (i) the KS potential of RPA-OEP is sufficiently accurate and (ii) other effects not included here such as electron-phonon coupling can be neglected. Then, the fact that the dielectric constants in the RPA are closer to experiment than the ones including local field effects can be entirely attributed to poor performance of the LDA kernel. In fact, it has been shown that the local and static nature of the LDA kernel has some serious physical flaws, see for example Ref. \onlinecite{Baerends2013}. This would further suggest that a more accurate kernel from MBPT would have to find once more a fine balance between vertex corrections in $\Sigma_{\rm xc}$ and $W$ to improve upon the RPA. It is in fact gratifying that the dielectric constants in the random-phase approximation calculated using the self-consistent RPA-OEP potential and orbitals agree on average so well with experiment. It indicates that the -- in itself closed -- RPA-OEP scheme is, considering its simplicity, an astoundingly accurate approximation working well beyond its expectations.
 
Finally, it is interesting that the QPA-RPA-OEP seems to yield somewhat better agreement with experiment than exact RPA-OEP, if the LDA exchange-correlation kernel is included. The effect of the QPA is strongest for ZnO and GaN, where exact RPA-OEP yields dielectric constants that are too large. Thus, the better agreement with experiment here can be understood as an error cancellation effect, since the band gaps are (incorrectly) larger in the QPA. That is, error cancellation can occur if both dynamical screening and many-body vertex corrections are neglected. A similar phenomenon, the so-called  ``$Z$-factor cancellation'',\cite{Kotani2007} has been previously discussed in the context of $GW$ calculations.

\section{\label{sec:Riemelmoser2021_6}Conclusion}

We have obtained optimized effective potentials from the random-phase approximation for a test set of 15 semiconductors and insulators. The main goal of the present work was to study the role of dynamical screening effects in the form of the quasiparticle approximation. We  have found that if this approximation is lifted, the band gaps are reduced towards PBE. Likely, this represents an improvement towards the true Kohn-Sham potential, as the RPA-OEP band gaps will be further reduced by the inclusion of ladder diagrams. 

As can be understood from a simple flat band model, the effect of the quasiparticle approximation is largest for materials with strong ionic character, where the proper treatment of the off-diagonal self-energy matrix elements is important. Indeed, our calculations have shown that for these materials (here ZnO, GaN and MgO) the differences in the OEP band gaps can amount to several 100 meV. As an interesting side result, we have found that the symmetric prescription [see Eq. \eqref{eq:Kotani}] for the off-diagonal elements in the quasiparticle approximation is likely not optimal.

We have also used the exchange-correlation potentials from the RPA-OEP methods to calculate macroscopic, ion-clamped dielectric constants. In agreement with the trend for the band gaps, the dielectric constants are generally underestimated by the QPA. The effect is, however, somewhat reduced here. For both RPA-OEP methods, the dielectric constants in the RPA show generally good agreement with experiment, much better than dielectric constants from PBE. Interestingly, we found that the inclusion of vertex corrections in form of local field effects is detrimental for most materials. We have concluded that the exchange-correlation kernel in the LDA is not sophisticated enough, though.  

Finally, from a technical point of view, we have demonstrated that the self-consistency cycle that is necessary for OEP can be sped up by pre-converging the exchange-correlation potential on a sparser $k$-point mesh. 

\section*{Acknowledgements}

Computation time at the Vienna Scientific Cluster (VSC) is gratefully acknowledged. 

\section*{Data availability}

The data that support the findings of this study are available from the corresponding author upon reasonable request.

\begin{appendix}

\section{$GW$ correlation energies for the HEG}\label{App:Riemelmoser2021_A}

In the following, we discuss briefly our numerical calculations of $GW$ correlation energies for the HEG. First, we describe the static QPA and COHSEX approximations and thereafter the fully self-consistent case.

\begin{figure}[!!tb]
\centering
\includegraphics [width=\linewidth,keepaspectratio=true] {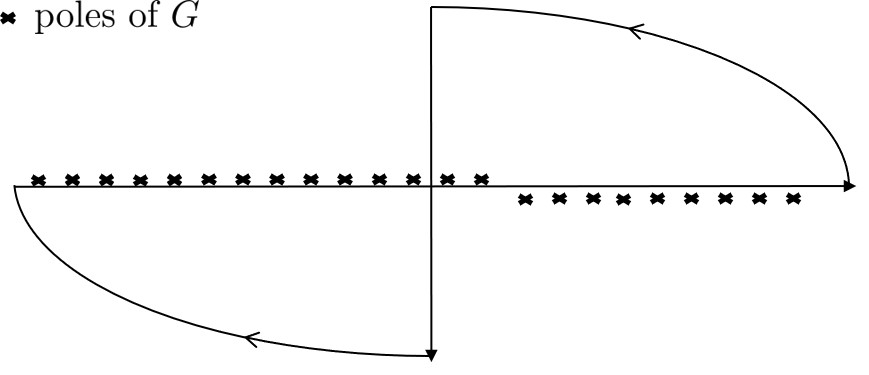}
\caption{Integration contour for the dynamical remainder term, compare Eq. \eqref{eq:dynamic_remainder}, for $\omega<\mu$. The integral over imaginary frequencies corresponds to $\Sigma^{\rm c}_{\rm xc}$, the poles of $G$ contribute to $\Sigma_{\rm xc}^{\rm p}$.}
\label{fig:contour}
\end{figure}

The $GW$ self-energy for the HEG is given by\cite{Hedin1970}
\begin{equation}\label{eq:GW_HEG}
\begin{aligned}
&\Sigma_{\rm xc,HEG}(\textbf{k},\omega) =\\ &-\uint \frac{\text{d}\nu}{2\pi i}\int \frac{\text{d}\textbf{q}}{(2\pi)^3} 	G(\textbf{k}+\textbf{q},\omega+\nu) W(\textbf{q},\nu) .
\end{aligned}
\end{equation}
As discussed in Sec. \ref{sec:model_QPA}, for static self-energies it is sufficient to consider the non-interacting $G_0W_0$ case, where the energy dispersion is given by $\varepsilon(k)=k^2/2$ and the chemical potential via $\mu = k_{\rm F}^2/2$ respectively. 	Following Hedin,\cite{Hedin1965} we split the screened interaction $W$ in a static term and a dynamical remainder term 
\begin{equation}
\begin{aligned}
&W(k,\nu) = \\
&\frac{V(k)}{\epsilon_{\rm HEG}^{\rm RPA}(k,0)}+V(k)\left[\frac{1}{\epsilon^{\rm RPA}_{\rm HEG}(k,\nu)}-\frac{1}{\epsilon^{\rm RPA}_{\rm HEG}(k,0)}\right] ,
\end{aligned}
\end{equation}
where $\epsilon^{\rm RPA}_{\rm HEG}=(1-\chi_{0,\rm HEG}V)^{-1}$ is the Lindhard dielectric function and it is understood that the static interaction with the polarization cloud, $W_{\rm p}(\omega)=W(\omega=0)-V$, is given by the average of infinitesimal positive and negative times\cite{Hedin1970}
\begin{equation}
W_{\rm p}(k,t)= W_{\rm p}(k)\left[\delta(t^{+})/2+\delta(t^{-})/2\right] .
\end{equation}
Then, performing the contour integral in Eq. \eqref{eq:GW_HEG} for the static parts yields the static Coulomb hole (COH) and screened exchange (SEX) terms, (see \citet{Hedin1965})
\begin{equation}
\begin{aligned}
\Sigma_{\rm xc,HEG}^{\rm COH}(k) =& \frac{1}{2}\int_0^{\infty}\frac{\text{d}q }{(2\pi)^3}4\pi q^2 V(q)\left[\frac{1}{\epsilon^{\rm RPA}_{\rm HEG}(q,0)}-1\right]\\
\Sigma_{\rm xc,HEG}^{\rm SEX}(k) =& -\int_0^{\infty} \frac{\text{d}q }{(2\pi)^3}2\pi q^2 \int_{-1}^1 \text{d}\xi  \\
&\times \frac{V(q)}{\epsilon^{\rm RPA}_{\rm HEG}(q,0)} \theta(k_{\rm F}^2-k^2-q^2-2kq\xi),
\end{aligned}
\end{equation}
The angular integral can be performed analytically using
\begin{equation}
\begin{aligned}
&\int_1^1 d\xi \theta(a-b\xi) \\ &=\left(\frac{a}{b}-1\right)\theta(b-a)-\left(\frac{a}{b}+1\right)\theta(-b-a)+2,
\end{aligned}
\end{equation}
thus both terms can be easily evaluated by numerical integration. The COH term is dispersionless, since the interaction with the static Coulomb hole is local in space. The SEX term reduces to the bare exchange result \eqref{eq:Sigma_EXX} for $\epsilon^{\rm RPA}_{\rm HEG}\rightarrow 1$. To evaluate the dynamical remainder term, one can deform the contour as shown in Fig. \ref{fig:contour}, yielding
\begin{equation}\label{eq:dynamic_remainder}
\begin{aligned}
&\Sigma^{\rm c}_{\rm xc,HEG}(k,\omega)= \\
&\frac{1}{2} \int_0^{\infty}\frac{\text{d}\nu}{2\pi}\int_0^{\infty} \frac{\text{d}q}{(2\pi)^3}2\pi q^2 \left[\frac{V(q)}{\epsilon^{\rm RPA}_{\rm HEG}(k,i\nu)}-\frac{V(q)}{\epsilon^{\rm RPA}_{\rm HEG}(q,0)}\right]\\
&\times\frac{1}{2kq}\frac{\ln(\omega-k^2/2-q^2/2-kq)^2+\nu^2}{\ln(\omega-k^2/2-q^2/2+kq)^2+\nu^2}\\
&\Sigma^{\rm p}_{\rm xc,HEG}(k,\omega) =  -\int_0^{\infty} \frac{\text{d}q }{(2\pi)^3}2\pi q^2 \int_{-1}^1 \text{d}\xi  \\
&\times \left[\frac{V(q)}{\epsilon^{\rm RPA}_{\rm HEG}(q,\omega)}-\frac{V(q)}{\epsilon^{\rm RPA}_{\rm HEG}(q,0)}\right] \\
&\times \left[\theta(2\omega-k^2-q^2-kq\xi)-\theta(k_{\rm F}^2-k^2-q^2-kq\xi)\right],
\end{aligned}
\end{equation}
where $\Sigma^{\rm c}_{\rm xc,HEG}$ involves an integral along the imaginary frequency axis and we have used that the integrand is an even function of $i\nu$. Furthermore, $\Sigma^{\rm p}_{\rm xc,HEG}$ corresponds to a contribution from the poles of $G$. The QPA is obtained by setting $\omega=k^2/2$ in Eq. \eqref{eq:dynamic_remainder}, hence
\begin{equation}
\begin{aligned}
\Sigma_{\rm xc,HEG}^{\rm QPA}(k) &= \Sigma_{\rm xc,HEG}^{\rm COH}(k)+\Sigma_{\rm xc,HEG}^{\rm SEX}(k)\\&+\Sigma_{\rm xc,HEG}^{\rm c}(k,k^2/2)+\Sigma_{\rm xc,HEG}^{\rm p}(k,k^2/2) .
\end{aligned}
\end{equation}
Again, these integrals can be evaluated numerically with relative ease. Our results for $\Sigma_{\rm xc,HEG}(k_{\rm F},k_{\rm F}^2/2)$ are in excellent agreement with those reported in Table III of Ref. \onlinecite{Hedin1965}. Finally, the $GW$ correlation energies for COHSEX and QPA are evaluated via Eq. \eqref{eq:GW_correlation_energy_static}. As was already discussed by Hedin,\cite{Hedin1965} the real part of $\Sigma^{\rm p}_{\rm xc,HEG}(k,k^2/2)$ vanishes at the Fermi edge and is also for other values of $k$ numerically very small. The difference between QPA and COHSEX therefore stems mostly from $\Sigma^{\rm c}_{\rm xc,HEG}$, which is positive. 

We now briefly discuss what changes in the case of self-consistent calculations. For the sake of clarity, we start with the $GW_0$ scheme, were only $G$ is iterated to self-consistency. 
For this purpose, it is useful to consider the Galitskii-Migdal energy in the form\cite{Holm1998}
\begin{equation}\label{eq:Galitskii_spectral}
\varepsilon_{\rm xc,HEG}^{\rm GM} = \frac{1}{n} \int_0^{\infty} \frac{\text{d}k}{(2\pi)^3} 4\pi k^2 \int_0^{\mu} \frac{\text{d}\omega}{2\pi}[\omega + \varepsilon(k)]A(k,\omega),
\end{equation}
where $A(k,\omega)$ is the spectral function. In any static approximation, it is given by $\delta[\omega-\varepsilon(k)-\Sigma_{\rm xc}(k)]$, hence we recover Eq. \eqref{eq:GW_correlation_energy_static}. In reality, some spectral weight is transferred to a plasmon satellite. Furthermore, the peaks are spread out, owing to the finite QP lifetime, which also implies a correlation part of the kinetic energy. However, it turns out that for the description of total energies only weight and position of the peaks, rather than their detailed structure, are of greater importance.\cite{Holm1998} Therefore, we can assume a sharp QP peak at $\varepsilon(k)+\Sigma_{\rm xc}(k)$ and a sharp plasmon peak at $\varepsilon(k)+\Sigma_{\rm xc}(k)-\omega_{\rm p}$, respectively, where $\omega_{\rm p}=\sqrt{4\pi n}$ is the plasma frequency. In this approximation, the spectral function is thus given by
\begin{equation}
\begin{aligned}
A(k,\omega) &\approx Z \delta[\omega-\varepsilon(k)-\Sigma_{\rm xc}]\\&+(1-Z)\{\delta[\omega-\varepsilon(k)-\Sigma_{\rm xc}+\omega_{\rm p}]\} .
\end{aligned}
\end{equation}
By inserting this result into Eq. \eqref{eq:Galitskii_spectral}, we find that in the QPA, the exchange-correlation energies are reduced by
\begin{equation}
\varepsilon_{\rm xc,HEG}^{\rm GM,QPA}- \varepsilon_{\rm xc,HEG}^{\rm GM} \approx  \frac{(1-Z) \omega_{\rm p}}{2} .
\end{equation}
Comparing with $GW_0$ data from \myciteauthor{Holm1998},\cite{Holm1998,Holm1999} we find that this simple formula is accurate to roughly 25\% for metallic densities.
In fully self-consistent $GW$ calculations, the plasmon peak is very much spread out, which may cast doubt on the assumptions above. However, it turns out that the effect on total energies is again small. 

For our self-consistent $GW$ calculations, we use the GM routines of \texttt{VASP} as described in Ref. \onlinecite{Kaltak2020}. We find that the self-consistency loop converges rather quickly, and 5 iterations are typically enough to converge the GM energies up to 1 mHa.
Such fully self-consistent $GW$ calculations for the HEG were performed previously by \myciteauthor{Holm1998} \cite{Holm1998,Holm1999} as well as \myciteauthor{GarciaGonzalez2001},\cite{GarciaGonzalez2001} and our exchange-correlation energies agree well with theirs (up to 1 mHa).
\section{Numerical solution of the linearized Sham-Schlüter equation}\label{App:Riemelmoser2021_B}
In this section we describe in detail our implementation of step (iv) in the RPA-OEP self-consistency scheme, compare Fig.
\ref{fig:self_consistency_scheme}.

In the $i$-th iteration of the self-consistency cycle, we determine the difference between the effective exchange-correlation
potential of the current and the previous iteration, $\Delta v_{\rm xc}^{(i)}=v_{\rm xc}^{(i)}-v_{\rm xc}^{(i-1)}$. Furthermore, the non-interacting Green's function of the previous iteration, $G_0^{(i-1)}(\omega)=[
\omega - H_{\rm KS}^{(i-1)}]^{-1}$, belongs to the Kohn-Sham Hamiltonian
$H_{\rm KS}^{(i-1)}=T+v_{\rm ext}+v_{\rm H}+v_{\rm xc}^{(i-1)}$. Then, $\Delta v_{\rm xc}^{(i)}$ is
obtained from the linearized SSE (\ref{eq:LSSE})
as follows. 
Writing Tr for the integration over $\mathbf{r}'$ and $\omega$, Eq. (\ref{eq:LSSE}) is rewritten into 
\begin{equation}
	0=\mathrm{Tr}\left\lbrace
	2 G_0^{(i-1)}\left[\Sigma_{\rm xc}^{(i-1)}-v_{\rm xc}^{(i-1)}
	\right]G_0^{(i-1)}\right\rbrace (\mathbf{r}).
\end{equation}
Add a zero of the form $0=H_{\rm KS}^{(i-1)}-H_{\rm KS}^{(i-1)}$ between the $G_0W_0$
self-energy and the exchange-correlation potential and separate the self-energy into an exchange
and correlation term, $\Sigma_{\rm xc}^{(i-1)}=\Sigma_{\rm x}^{(i-1)}+\Sigma_{\rm c}^{(i-1)}$. This yields
\begin{equation}
	\label{eq:Naturalo}
	\begin{split}
		&\mathrm{Tr}\left\lbrace\chi_0^{(i-1)}\Delta
		v_{\rm xc}^{(i)}\right\rbrace(\mathbf{r})=\\&\mathrm{Tr} \left\lbrace 2 \left[H_{\rm HF}^{(i-1)}-H_{\rm KS}^{(i-1)}
		\right]G_0^{(i-1)}G_0^{(i-1)}\right\rbrace(\mathbf{r})\\
		+&\mathrm{Tr}\left\lbrace
	2 G_0^{(i-1)}\Sigma_{\rm c}^{(i-1)}G_0^{(i-1)}\right\rbrace(\mathbf{r}).
	\end{split}
\end{equation}
The right hand side is evaluated in the basis of the Hartree-Fock
Hamiltonian $H_{\rm HF}^{(i-1)}=T+v_{\rm ext}+v_{\rm H}+v_{\rm x}^{(i-1)}$ and diagonalized,
yielding natural orbitals $\phi_\alpha^{(i-1)}$ and occupancies $f_\alpha^{(i-1)}$ for the
$i$-th iteration. The corresponding natural orbital charge density, 
\begin{equation}
	\Delta \rho_{\rm xc}^{(i-1)}(\mathbf{r}) = \sum\limits_\alpha^{\rm all} f_\alpha^{(i-1)}
	\phi_\alpha^{*(i-1)}(\mathbf{r}) \phi_\alpha^{(i-1)}(\mathbf{r}),
\end{equation}
is a valid spectral representation of the right hand side of Eq. (\ref{eq:Naturalo})
such that the linearized SSE reduces to
\begin{equation}
	\int \mathrm{d}\mathbf{r}'\chi_0^{(i-1)}(\mathbf{r},\mathbf{r}')\Delta
                v_{\rm xc}^{(i)}(\mathbf{r}')=\Delta\rho_{\rm xc}^{(i-1)}(\mathbf{r}) .
\end{equation}
This equation is Fourier transformed to reciprocal space 
and solved for $\Delta v_{\rm xc}^{(i)}$. 
We follow Ref. \onlinecite{Bleiziffer2013} and use a cutoff $t_{\rm SVD}=10^{-4}$ in the inversion of $\chi_0$, which eliminates eigenmodes with eigenvalues smaller than $t_{\rm SVD}$.
Updating the Kohn-Sham Hamiltonian via $H_{\rm KS}^{(i)}=H_{\rm KS}^{(i-1)}+\Delta
v_{\rm xc}^{(i)}$ completes one self-consistency loop. 

\end{appendix}
\bibliography{Riemelmoser2021_postprint}
\printindex
\end{document}